\documentclass[%
 aip,
 amsmath,amssymb,
 reprint,%
]{revtex4-1}

\usepackage{graphicx}
\usepackage{dcolumn}
\usepackage{bm}

\usepackage[utf8]{inputenc}
\usepackage[T1]{fontenc}
\usepackage{mathptmx}
\usepackage{etoolbox}

\usepackage{algorithm}
\usepackage{algorithmic}
\usepackage{placeins}

\DeclareMathOperator*{\argmin}{arg\,min}
\usepackage{multirow}
\usepackage[figuresleft]{rotating}
\usepackage{subcaption}
\usepackage{CJKutf8}

\usepackage{color}
\usepackage{soul}

\makeatletter
\def\@email#1#2{%
 \endgroup
 \patchcmd{\titleblock@produce}
  {\frontmatter@RRAPformat}
  {\frontmatter@RRAPformat{\produce@RRAP{*#1\href{mailto:#2}{#2}}}\frontmatter@RRAPformat}
  {}{}
}%
\makeatother
\begin{document}

\preprint{AIP/123-QED}

\title{Phase proper orthogonal decomposition of non-stationary turbulent flow}
\author{Yisheng Zhang (\begin{CJK*}{UTF8}{gbsn}张益盛\end{CJK*})}
 \email{yiszh@dtu.dk.}
\affiliation{Department of Civil and Mechanical Engineering, Technical University of Denmark (DTU), Kgs. Lyngby, 2800, Denmark}
\author{Azur Hod\v zi\'c}
\affiliation{Department of Civil and Mechanical Engineering, Technical University of Denmark (DTU), Kgs. Lyngby, 2800, Denmark}
\author{Fabien Evrard}
 \affiliation{Institute for Process Engineering, Otto-Von-Guericke-University Magdeburg, Magdeburg, Saxony-Anhalt, 39106, Germany}
 \author{Berend Van Wachem}
 \affiliation{Institute for Process Engineering, Otto-Von-Guericke-University Magdeburg, Magdeburg, Saxony-Anhalt, 39106, Germany}
\author{Clara M. Velte}
 \affiliation{Department of Civil and Mechanical Engineering, Technical University of Denmark (DTU), Kgs. Lyngby, 2800, Denmark}
 
\date{\today}

\begin{abstract}
A phase proper orthogonal decomposition (Phase POD) method is demonstrated, utilizing phase averaging for the decomposition of spatio-temporal behaviour of statistically non-stationary turbulent flows in an optimized manner. The proposed Phase POD method is herein applied to a periodically forced statistically non-stationary lid-driven cavity flow, implemented using the snapshot proper orthogonal decomposition algorithm. Space-phase modes are extracted to describe the dynamics of the chaotic flow, in which four central flow patterns are identified for describing the evolution of the energetic structures as a function of phase. The modal building blocks of the energy transport equation are demonstrated as a function of the phase. The triadic interaction term can here be interpreted as the convective transport of bi-modal interactions. Non-local energy transfer is observed as a result of the non-stationarity of the dynamical processes inducing triadic interactions spanning across a wide range of mode numbers.
\end{abstract}

\maketitle


\section{Introduction}\label{sec:introduce}

Analysis of non-stationary turbulent flow dynamics is characterised by highly complex features, exhibiting richer interactions between coherent structures than statistically stationary turbulent flows. Given the large amounts of data available from computational fluid dynamics (CFD) simulations and comprehensive experimental measurement techniques, such as time-resolved particle image velocimetry (TR-PIV), and four-dimensional particle tracking velocimetry (4D-PTV), the emergence of statistically non-stationary turbulence measurements can be expected to become ever more prevalent in the future \citep{Rosi2018, Lacassagne2020,Steiros2022}. Given that statistically non-stationarity turbulent flows are more likely to occur in nature and in industrial processes than stationary flows, the present work demonstrates a decomposition method for non-stationary turbulent/chaotic flows based on the proper orthogonal decomposition (POD). By including the temporal (or phase) domain into the domain of the POD eigenfunctions, the POD can be used to analyze dynamic coherent structures that are functions of  space and time for statistically non-stationary turbulent flows, and thereby be used to analyse flows that cannot be expected to follow the classical turbulence theory of stationary flows \citep{Kolmogorov1941a,Kolmogorov1941b,Kolmogorov1941c,Batchelor1982}.

The POD was introduced to the fluid mechanic community by Lumley in a generalized spatio-temporal domain, facilitated by an ensemble averaging operator for the generation of the POD kernel\cite{Lumley1967, Lumley1972}. Nevertheless, most of the POD implementations were historically performed using the time-averaging operation applied to statistically stationary flows, producing POD eigenfunctions defined exclusively on a spatial domain. This formulation of the POD is popularly termed \textit{space POD}. These spatial POD modes represent an optimal basis $\lbrace\varphi^{\alpha}(\bm{x})\rbrace_{\alpha=1}^\infty$ in a Hilbert space $\mathcal{H}=\overline{\mathrm{span}}\{ \varphi^\alpha \}_{\alpha=1}^\infty$, where any realization $u(\bm{x},t)$ can be expressed as a sum of the products of temporal projection coefficients and the POD modes

\begin{equation} \label{eq:ubase}
u(\bm{x},t)=\sum^\infty_{\alpha=1} a_\alpha(t) \varphi^{\alpha}(\bm{x}).
\end{equation}

By solving the POD eigenvalue problem, an optimal basis, $\lbrace\varphi^{\alpha}(\bm{x})\rbrace_{\alpha=1}^N$, is obtained, which is maximally robust against truncation with respect to a Hilbert space norm. However, the lack of temporal information characterizing the space POD is limiting in addition to its restriction to statistically stationary flows.

Efforts to extend the implementation from space POD to the space-time POD or the space-frequency POD have been contributed by researchers since the 1980s. Glauser and George\cite{GlauserGeorge1987}, George\cite{George1988} and Glauser et al.\cite{GlauserLeibGeorge1987} applied the Fourier Transform to stationary and homogeneous coordinates in an axisymmetric jet mixing layer experiment to capture the dynamics of turbulence. In this view, the new POD eigenfunctions include space-frequency information, and the coefficients indicate turning the POD modes on and off at various points across the temporal evolution. Using 138 simultaneously sampled hotwire anemometer probes in a high Reynolds number axisymmetric shear layer experiment, Citriniti and George\cite{Citriniti2000} analysed the production mechanism and instability transportation of spatial structures with the dynamics of low-order POD modes. Expanding the experiment by Citriniti and George\cite{Citriniti2000}, Jung et al.\cite{Jung2004} and Gamard et al.\cite{Gamard2004} studied the evolution of the most energetic features in both the near-field region and the far field region with the POD eigenspectra (the double Fourier transform in time and azimuthal direction of the correlation tensor) at jet exit Reynolds numbers ranging from 40\, 000 to 156\, 800. Towne et al.\cite{Towne2018} and Schmidt and Colonius\cite{Schmidt2020} proposed the spectral proper orthogonal decomposition (SPOD) by applying the Fourier transform to the covariance tensor to compute eigenfunctions in stationary turbulent flows. Coherent in both space and time, the SPOD modes can reveal the underlying flow physics, such as the presence of vortex shedding, flow instabilities, acoustic resonances, and so on. Muralidhar et al.\cite{Muralidhar2019} performed a full spatio-temporal POD of a channel flow under the condition of stationarity (again using a Fourier series decomposition in time). The SPOD can therefore capture the four-dimensional spatio-temporal coherent structures of turbulent flows but is inherently restricted to stationary flows. Note that some researchers take the Hilbert space $L^2([0,+\infty))$ or $L^2((-\infty,+\infty))$ as the default space when they transform the temporal domain into the frequency domain in stationary flows, even though the Fourier modes are not a basis in the Hilbert space $L^2([0,+\infty))$ or $L^2((-\infty,+\infty))$ due to non-square-integrability in domain $[0, +\infty)$ or $(-\infty, +\infty)$ \citep{Hodzic2022}. 

In the scientifically and technologically important case of non-stationary flows, however, all the POD methods above fall short. To capture the dynamics of non-stationary flows, Gordeyev and Thomas\cite{Gordeyev2013} proposed a temporal POD (TPOD) technique by retaining the temporal information of the flow in both the eigenvalues as well as the eigenfunctions, of which the low-dimensional models correctly reflect the transient behaviour from initial flow states. The transient behaviour of the temporal POD modes is produced by editing space modes as a snapshot in time series. The four-dimensional spatio-temperal coherent structures are not resolved in the TPOD. Schmidt and Schmid\cite{Schmidt2019} proposed a conditional averaging operation in combination with a space-time POD in an example of studying the intermittent acoustic radiation of a turbulent jet flow. However, one common characteristic resulting from the inclusions of temporal information in the aforementioned examples, is their deviation from an analogous POD formulation implemented using an ensemble averaging on statistically non-stationary flow fields (see p. 2 in Ref.~\onlinecite{Lumley1967}). The statistical approach to the theory of turbulence is based on the consideration of the statistical ensemble of all similar flows created by some set of fixed external conditions, and the ensemble averaging represents the theoretical mean over all possible states of the system (see p. 209 and p. 215 in Ref.~\onlinecite{Monin1971}).

The applications of POD and POD-extended techniques are active in recent studies of stationary flows. Carlson et al. applied POD and the calibration of the model with test data to aeroservoelastic model of an aircraft for reducing the differences between computational model-based predictions and flight-test measurements \cite{Carlson2017}. Riches et al. utilize POD to analyze the wake-dynamics of a low-mass ratio circular cylinder \cite{Riches2018}. Charalampous et al. explore the primary atomization in a coaxial stream of high-speed gas and liquid by POD with different gas-to-liquid momentum ratios \cite{Charalampous2019}. He et al. applied SPOD on the near-tip flow field and the blade surface pressure of a low-speed compressor rotor to identify the spatiotemporal coherent structure of compressor tip leakage flow \cite{He2021}. Blanc et al. use POD to investigate coherent structures revealed by the thermal field in the molten pool during gas tungsten arc welding on thin plates \cite{Blanc2021}. The internal unsteady flow structure in a centrifugal pump is studied with POD for evaluating the performance of the centrifugal pump by Zhang et al.\cite{Zhang2021}. Schi{\o}dt et al. demonstrated particle proper orthogonal decomposition for extraction of temporal statistical information on dispersed (discrete) phases of multiphase flows \cite{Schioedt2022}. Olesen et al. presented a dissipation-optimized POD for spanning a wide range of spatial scales in a turbulent channel flow\cite{Olesen2023}. A POD technique for non-stationary flows is needed to analyze the four-dimensional spatiotemporal coherent structures.

In the work of Berkooz et al.\cite{Berkooz1993}, the application of phase averaging in combination with the POD was proposed in order to retain the optimality criterion while capturing the temporal dynamics of the field. The present work demonstrates a phase proper orthogonal decomposition (Phase POD) method, making use of the proposed phase averaging, which can readily be applied for non-stationary turbulent flow fields exposed to periodic forcing. The temporal coordinate is here treated analogously to the inhomogeneous spatial coordinates, which means that the domain of the eigenfunctions is the Cartesian product of the spatial and phase domain. This implies that the projection coefficients obtained by projecting a realisation of the statistically non-stationary flow field onto the eigenfunctions obtained from the Phase POD are constants (independent of time) as opposed to those appearing in Galerkin projected transport equations used to analyse stationary energy transfer \citep{Couplet2003}, in reduced-order models based on space POD \citep{Balajewicz2013, Lee2020}, as well as in the sparsification methods for energy interaction analysis \citep{Rubini2020, Rubini2022}.

This paper aims to demonstrate the Phase POD technique and its application to statistically non-stationary turbulent flow. In addition to the formalism of the Phase POD, the key points we investigate are:
\begin{enumerate}
\item The spatio-temporal structure contained in the Phase POD modes.
\item The efficiency of the Phase POD algorithm.
\item The unique information that the Phase POD is able to extract for the case of non-stationary turbulent flows.
\end{enumerate}

Section \ref{sec:PhasePOD} formalizes the proposed Phase POD including general forms, numerical implementation and computationally effective algorithm of the Phase POD based on the snapshot POD method. Section \ref{sec:example} presents the application of the Phase POD to a statistically non-stationary lid-driven cavity flow, and the analysis of energy budget and triadic interactions with the Phase POD modes. Non-local energy transfer is observed in subsection \ref{sec:non-local} as a result of the non-stationarity of the dynamical processes.


\section{Phase POD}\label{sec:PhasePOD}
Let $\mathcal{H}$ be a Hilbert space with inner product $(\cdot, \cdot):\mathcal{H}\times\mathcal{H}\rightarrow \mathbb{C}$ and norm, $\Vert \cdot \Vert:\mathcal{H}\rightarrow \mathbb{R}_{+}$, defined as

\begin{equation}
 \Vert \bm{f} \Vert=\sqrt{(\bm{f},\bm{f})}\,,\quad\bm{f}\in\mathcal{H}\,.
\end{equation}

Consider a set of $N$ velocity realizations $\{ \bm{u}^\beta \}_{\beta=1}^N\subseteq \mathcal{H}$. The POD seeks a set of basis functions that optimally expand the set of velocity realizations with respect to some optimization criterion. In its application to fluid mechanical data, the optimization criterion is often chosen to be the total turbulent kinetic energy (TKE) of the field (see Chapter 3 in Ref.~ \onlinecite{Holmes2012}). In this paper, index notation follows tensor calculus notation; superscripts $\cdot^{\alpha, \beta, \gamma}$ refer to contravariant vectors, and subscripts $\cdot_{\alpha, \beta, \gamma}$ refer to covariant vectors; $i, j$ denote velocity and coordinate system components. The mathematical formulation of the optimality problem is in this case expressed as the minimization of the residual norm between $\{ \bm{u}^\beta \}_{\beta=1}^N$ and its projection onto $\lbrace\bm{\varphi}^{\alpha}\rbrace_{\alpha=1}^N\subset \mathcal{H}$ \cite{Holmes2012}, 
\begin{equation} \label{eq:PODstart}
\argmin_{\lbrace\bm{\varphi}^{\alpha}\rbrace_{\alpha=1}^N\in\mathcal{H}}\Bigl\langle\Bigl\lbrace\Bigl\Vert \bm{u}^\beta-\frac{(\bm{u}^\beta,\bm{\varphi}^{\alpha})}{\Vert\bm{\varphi}^{\alpha}\Vert^2}\bm{\varphi}^{\alpha}\Bigr\Vert^2\Bigr\rbrace_{\beta=1}^N\Bigr\rangle\,,
\end{equation}
where $\langle \cdot \rangle$ is an averaging operator taken over all data realizations, where the vector fields, $\{ \bm{u}^\beta \}_{\beta=1}^N$, are considered to be independent random variables.  The search for the optimal basis then reduces to solving the eigenvalue problem for the operator $\mathcal{R}:\mathcal{H}\rightarrow\mathcal{H}$,
\begin{equation} \label{eq:PODend}
\mathcal{R}\bm{\varphi}^{\alpha}=\lambda_{\alpha}\bm{\varphi}^{\alpha}\,,
\end{equation}
where $\mathcal{R}$ is defined as $\mathcal{R}\bm{\varphi}^{\alpha}=\langle\lbrace(\bm{\varphi}^{\alpha},\bm{u}^{\beta})\bm{u}^{\beta}\rbrace_{\beta=1}^N\rangle$ and the eigenfunctions, $\{ \bm{\varphi}^\alpha \}_{\alpha=1}^N $, of the operator $\mathcal{R}$ are the set of POD modes satisfying $\overline{\text{span}}\{ \bm{\varphi}^\alpha \}_{\alpha=1}^N=\mathcal{H}$. All velocity realizations can therefore be spanned as
\[
\bm{u}^\beta = a_\alpha^\beta\bm{\varphi}^\alpha, \quad a_\alpha^\beta=(\bm{u}^\beta, \bm{\varphi}^\alpha)\,,
\]
here the Einstein notation applies and $\alpha$ is the dummy index. The projection coefficients are uncorrelated, satisfying
\[
\bigl\langle\bigl\lbrace a_\beta^{\alpha*} a_{\alpha'}^{\beta} \bigr\rbrace_{\beta=1}^N\bigr\rangle=\lambda_\alpha\delta_{\alpha'}^{\alpha}\,,
\]
where $\delta_{\alpha'}^{\alpha}$ is the Kronecker delta, and $^*$ denotes the complex-conjugation operator. More general details on the POD can be found in Refs.~\onlinecite{Holmes2012, Berkooz1993, Sirovich1987}.

Ensemble averaging can be viewed as the most general averaging operation applied in dynamical systems and statistical theory (see Refs.~\onlinecite{Frisch1995, Monin1971}). It is in principle computed from data sets obtained from a series of independent experiments. Each member of the ensemble will have nominally identical boundary conditions and fluid properties and members of an ensemble are statistically independent. Although ensemble averaging may be considered as the most general averaging operation conserving the spatio-temporal domain of the eigenfunctions, it is especially useful for the application of the POD on statistically non-stationary flows, in order to modally characterize the spatio-temporal dynamics. However, for the class of periodically forced, statistically non-stationary flows, ensemble averaging can be considered to be equivalent to phase averaging (see Appendix \ref{appA}). In these cases, the eigenfunctions of the phase averaged POD or Phase POD, contain the information of the spatio-temporal coherent structures and the eigenvalues denote the statistical energy contributions of these eigenfunctions for the reconstruction of the flow. The Phase POD is therefore constructed by an incorporation of a spatio-temporal domain with a phase averaging operation in the definition of the POD operator. Although this limits its use to cases where phase averaging is a meaningful averaging operation to employ, such as cases of periodically forced flows or transient flows that are reproducible, 1) these cases often appear in industry processes and 2) periodically forced cases may be used to model more general non-stationary processes occurring in nature which may be considered as quasi-periodical. The aim is then for the Phase POD to allow us to study statistically non-stationary chaotic/turbulent fields in an optimal way, where the modes are a function of space and phase, thereby exhibiting the spatio-temporal coherence that is often sought in the analysis of physically observable coherent structures. The phase averaging operation used in the current work differentiates itself from the one defined in {\it Statistical Mechanics} (see p. 186 in Ref.~\onlinecite{Toda1983}) and {\it Ergodic theory} (see p. 2 in Ref.~\onlinecite{Parry1981}), which describe a possible state in the phase space. Instead, the phase concept throughout this paper exclusively denotes the time steps within each period.

Let $\Omega_S$ denote the spatial domain of the velocity field. Then a space-time domain can be defined as ${\Omega}:=\Omega_S \times \Omega_T$, $\Omega_T=(0,T]\subset\mathbb{R}_+$, such that the Hilbert space of interest is 
\begin{equation}
    \mathcal{H}:=\left\lbrace{\bm{f}}:{\Omega}\rightarrow \mathbb{C}^m\, \left\vert\, \Vert {\bm{f}} \Vert <\infty \right.\right\rbrace\,
\end{equation}
where $m\in\mathbb{N}_0$ denotes the number of dimensions of the vector field (e.g. for $m=1$ the field $\bm{f}$ represents a scalar field and for $m=3$ it represents a three-dimensional vector field).
With $\mathcal{H}$ as the domain of the POD operator, the eigenfunctions of Eq.~\eqref{eq:PODend} represent the energetically optimal spatio-temporal basis for $\mathcal{H}$. Since the domain of the eigenfunctions in Eq.~\eqref{eq:PODend} is a spatio-temporal domain, the resulting formulation of $\mathcal{R}$ corresponds to the original definition introduced by Lumley\cite{Lumley1967}, differing by the phase averaging operation for the generation of the POD kernel. The phase averaging operation, $\langle \cdot \rangle:\mathcal{K}\rightarrow \mathcal{H}$, where
\begin{equation}
    \mathcal{K}:=\left\lbrace{\bm{g}}:{\Omega_S\times\mathbb{R}_+}\rightarrow \mathbb{C}^m \, \left\lvert \,\frac{1}{N_T}\sum^{N_T}_{n=1}\left\vert\bm{g}(\bm{r},p+nT)\right\vert<\infty \right.\right\rbrace\,,
\end{equation}
is defined as
\begin{equation} \label{eq:phaseaverage}
\langle \bm{u} \rangle= \frac{1}{N_T}\sum^{N_T}_{n=1}\bm{u}(\bm{r},p+nT),
\end{equation}
where $\bm{r}\in \Omega_S$ is the spatial position, $n$ is the period number, $T$ is the period length, $N_T$ is the total number of periods, and $p\in [0,T]$ denotes the discrete or continuous phase parameter in each period. 
\subsection{Numerical implementation using classical POD}
In the discrete case, each period has the same number of phases and the total phase number, $N_p$, is chosen such that it satisfies the Nyquist criterion for resolving the time scale of interest. The sampling process extends over $N_T$ periods and resolves $N_p$ phases in each period, so the total temporal sampling dimension is $N_T \times N_p$. Each realization can thus be written as a function of $p$, $n$ and $T$, from which the discrete temporal variable of the underlying flow field is expressed as $t=p+nT$.

Let the discretely sampled velocity fields in a spatio-temporal domain be in
\begin{equation}
 \mathcal{H}:=\left\lbrace \bm{u}_i\in \mathbb{R}^{N_d}\, \left\vert\, \sum_{i=1}^{N_d}\left\vert \bm{u}_i\right\vert^2 <\infty \right.\right\rbrace\,,
\end{equation}
equipped with the inner product $\left(\cdot,\cdot\right):\ \mathcal{H}\times \mathcal{H}\rightarrow \mathbb{R}$, defined as
\begin{equation}
\left(\bm{u}^\alpha,\bm{u}^\beta\right)=\bm{u}^\alpha_i\cdot \bm{u}^{\beta *}_i\,,\quad\alpha,\beta\in\mathbb{N}\,,
\label{eq:innerproduct}
\end{equation}
where $i$ is the dummy index from 1 to $N_d$, the short notation $\bm{u^\alpha}=\left\lbrace\bm{u^\alpha_i}\right\rbrace_{i=1}^{N_d}$ is used to designate the $\alpha$-th velocity realization, $N_d$ is the total number of spatio-temporal degrees of freedom, and $(\cdot)$ designates the complex dot product over the $m$ velocity components. The empirical velocity data are acquired from a finite number of discrete spatio-temporal points and are structured in the form of a multi-dimensional matrix. Two-dimensional unsteady flow velocities are structured in a three-dimensional matrix, consisting of two spatial dimensions and one temporal dimension. Correspondingly, a three-dimensional unsteady flow is structured in a four-dimensional matrix. This subsection demonstrates formalism of three-dimensional unsteady flows, whereas Appendix \ref{appB} details the formalism for two-dimensional unsteady flows, which is applied to the statistically non-stationary cavity flow in section \ref{sec:example}. 

The velocity $\bm{u}^\alpha$ is in the Hilbert space $ \mathcal{H}$, including three velocity components $\{\bm{u}^{\alpha}_1, \bm{u}^{\alpha}_2, \bm{u}^{\alpha}_3 \}$. An inner product based on discrete data is defined from the definition of TKE as $(\bm{u}^\alpha,\bm{u}^\beta)=\bm{u}^\alpha_1\cdot \bm{u}^{\beta *}_1+\bm{u}^\alpha_2\cdot \bm{u}^{\beta *}_2+\bm{u}^\alpha_3\cdot \bm{u}^{\beta *}_3$.

For the sake of computation, we arrange the multi-dimensional velocity matrix $\bm{u}^\alpha$ to a column vector $\bm{q}^\alpha = \mathrm{vec}(\bm{u}^\alpha)$, and the algorithm is outlined as follows:

\begin{algorithm}[H]
\caption{Data vector arrangement with $N_d=3$}
\label{alg:a}
\begin{algorithmic}
\REQUIRE  $u_1^{\alpha}(\bm{x})$, $u_2^{\alpha}(\bm{x})$, $u_3^{\alpha}(\bm{x})$
\ENSURE $\bm{q}^\alpha=[\bm{q}^{\alpha}_1, \bm{q}^{\alpha}_2, \bm{q}^{\alpha}_3]^T$
\FOR{$p=1:N_p$}
    \FOR{$x_3=1:N_3$} 
        \FOR{$x_2=1:N_2$} 
            \FOR{$x_1=1:N_1$}
                \STATE $\bm{q}^{\alpha}_1(b)\gets u_1^{\alpha}(x_1,x_2,x_3,p)$
                \STATE $\bm{q}^{\alpha}_2(b)\gets u_2^{\alpha}(x_1,x_2,x_3,p)$
                \STATE $\bm{q}^{\alpha}_3(b)\gets u_3^{\alpha}(x_1,x_2,x_3,p)$
                \STATE $b\gets b+1$
            \ENDFOR
        \ENDFOR
    \ENDFOR
\ENDFOR
\end{algorithmic}
\end{algorithm}

Here $N_1$, $N_2$, $N_3$ and $N_p$ are discrete sampling points in the coordinates $x_1$, $x_2$, $x_3$ and $p$. The empirical velocity vector $\bm{q}^\alpha$ is a vector with $3 N_1  N_2 N_3 N_p$ rows. We take $M=3 N_1 N_2 N_3 N_p$ to simplify the expression of the data vector. Thus the empirical velocity vector $\bm{q}^\alpha$ is an vector with $M$ rows, and all the empirical velocity vectors can form a velocity matrix $\mathbf{u}$ with size of $M\times N_T$.
\begin{equation}
    \mathbf{u}=\left[
  \bm{q}_{1},\bm{q}_{2},\cdots,\bm{q}_{N_T}
  \right].
\label{eq:u}
\end{equation}

In vector form, the inner product in Eq.~\eqref{eq:innerproduct} can be written as vector dot product:
\[
(\bm{u}^{\alpha}, \bm{u}^{\beta}) = \bm{q}^{\alpha} \cdot \bm{q}^{\beta *}.
\]

In this Hilbert space $\mathcal{H}$, the eigenvalue problem Eq.~\eqref{eq:PODend} with empirical data becomes:
\begin{equation} \label{eq:PhasePODIntegral}
\mathcal{R}\bm{\varphi}^\alpha =  \langle\bm{q}^{\beta*}\bm{q}^{\beta}\rangle \bm{\varphi}^\alpha = \mathbf{R}\bm{\varphi}^\alpha = \lambda_\alpha \bm{\varphi}^\alpha,
\end{equation} 
where $\beta$ is the dummy index from 1 to $N_T$, $\mathbf{R}$ is the correlation matrix, $\mathbf{R}=\langle \lbrace\bm{q}^{\beta} \bm{q}^{\beta*}\rbrace_{\beta=1}^{N_T}\rangle$. In vector form, the correlation matrix is expressed as
\begin{equation}\label{eq:correlation}
\mathbf{R}=\langle\bm{q}^{\beta*}\bm{q}^{\beta}\rangle =
\begin{bmatrix}
\mathbf{R}_{11} & \mathbf{R}_{12} & \mathbf{R}_{13}\\
\mathbf{R}_{21} & \mathbf{R}_{22} & \mathbf{R}_{23}\\
\mathbf{R}_{31} & \mathbf{R}_{32} & \mathbf{R}_{33}\\
\end{bmatrix} ,
\end{equation}
where $\beta$ is a dummy index, $\mathbf{R}_{ij}$ is the correlation matrix of the $i$ component velocity and the $j$ component velocity, and $(i, j)$ means velocity components in $\{x_1, x_2, x_3\}$. The velocity data vector $\bm{q}^\beta$ in Eq.~\eqref{eq:u} denotes the rearranged velocity of the $\beta$th snapshot. So the correlation matrix in Eq.~\eqref{eq:correlation} can be written as
\begin{equation}
    \mathbf{R}=\langle\bm{q}^{\beta*}\bm{q}^{\beta}\rangle = \frac{1}{N_T}\sum^{N_T}_{\alpha=1}\bm{q}^{\beta *}\bm{q}^{\beta }= \frac{1}{N_T}\mathbf{u} \mathbf{u}^*.
    \label{eq:RU}
\end{equation}
The averaging method in the correlation matrix is phase averaging, and the velocity vectors are independent random variables. Due to the non-stationarity of the flow, time averaging cannot be applied to Eq.~\eqref{eq:PODend} and Eq.~\eqref{eq:PhasePODIntegral} because the ansatz of equivalence of time average and ensemble average cannot hold. We discuss the equivalence of phase averaging and ensemble averaging for periodically driven unsteady flow in Appendix \ref{appA}.
\subsection{Numerical implementation using the method of snapshots}\label{sec:computation}
The correlation matrix in Eq.~\eqref{eq:correlation} is an $M\times M$ matrix. If the four-dimensional velocity vector is sampled with 100 elements in each coordinate direction, the size of the correlation matrix will be $(3\times10^8)\times (3\times10^8)$, which will require in excess of $650\,\mathrm{Petabytes}$ (in 64-bit floating-point arithmetics). Present computers cannot handle such a large matrix. Hence, the snapshots method is recommended to process the eigenvalue problem. 

Sirovich\cite{Sirovich1987} introduced the snapshot POD method to handle the memory-intensive problem of space POD in the case $M\gg N_T$. Typically, the empirical data converges within several thousand data sets (or periods) according to the central limit theorem (see p. 342 in Ref.~\onlinecite{Hogg2019}). If the correlation matrix of the Phase POD with a size of $M\times M$ can be converted to a matrix with a size of $N_T\times N_T$ as the snapshot method of space POD does, the Phase POD can decompose all kinds of high-resolution data efficiently.

Assuming the empirical data converges within $N_T$ sets (or periods), the solution of the Phase POD from the correlation matrix $\mathbf{R}$ is
\begin{equation}
    \mathbf{R}\Phi = \Phi \Lambda,
    \label{eq:snapshots1}
\end{equation}
where the eigenfunction matrix $\Phi$ is formed from the eigenvectors,
\[
\Phi = [\bm{\phi}^1,\bm{\phi}^2,\cdots,\bm{\phi}^\alpha, \cdots,\bm{\phi}^{N_T},\bm{\phi}^{N_T+1}, \cdots,\bm{\phi}^{M}].
\]
Because the correlation matrix is computed from $N_T$ independent data vectors, the eigenvalue matrix is a rank $N_T$ matrix, written as $\Lambda =\mathrm{diag}[\lambda_1 , \lambda_2, \cdots ,\lambda_{N_T} , 0_{{N_T}+1},\cdots,0_M]$, where $\lambda_\alpha$ is the eigenvalue with the order $\lambda_1\ge\lambda_2\ge\cdots\lambda_{N_T}>0$. Thus, the eigenvectors $\{ \bm{\phi}^1,\bm{\phi}^2,\cdots,\bm{\phi}^{N_T} \}$ are deterministic from the empirical data,
\[
    \mathbf{R}\bm{\phi}^{\alpha} = \lambda_\alpha\bm{\phi}^{\alpha}, \alpha\in\{ 1, 2, \cdots,{N_T}\}.
\]

To solve Eq.~\eqref{eq:snapshots1}, we multiply Eq.~\eqref{eq:RU} by the empirical data matrix $\mathbf{u}$ defined in Eq.~\eqref{eq:u},
\begin{equation}\label{eq:snapshots2}
\mathbf{R}\mathbf{u} =\frac{1}{N_T}\mathbf{u} \mathbf{u}^* \mathbf{u} = \mathbf{u} (\frac{1}{N_T}\mathbf{u}^*\mathbf{u})= \mathbf{u}\mathbf{C}
\end{equation}
where $\mathbf{C}=\frac{1}{N_T}\mathbf{u}^*\mathbf{u}$. $\mathbf{C}$ is Hermitian, and can be eigendecomposited into $\mathbf{C}=\mathbf{QDQ}^*$ with unitary matrix $\mathbf{Q}$ and diagonal matrix $\mathbf{D}$. The diagonal elements of matrix $\mathbf{D}$ are the eigenvalues of matrix $\mathbf{C}$. So defining a new matrix $\Psi=\mathbf{u}\mathbf{Q}$, equation Eq.~\eqref{eq:snapshots2} is rewritten as
\begin{equation}
    \mathbf{R}\Psi=\mathbf{R}\mathbf{u}\mathbf{Q}= \mathbf{u}\mathbf{C}\mathbf{Q}= \Psi \mathbf{D}.
    \label{eq:snapshots3}
\end{equation}
The difference between Eq.~\eqref{eq:snapshots1} and Eq.~\eqref{eq:snapshots3} lies in the dimension of the matrices. In Eq.~\eqref{eq:snapshots3}, $\mathbf{D}$ is an $N_T\times N_T$ matrix and $\Psi$ is an $M\times N_T$ matrix. But in Eq.~\eqref{eq:snapshots1}, both $\Lambda$ and $\Phi$ are $M\times M$ matrices. Eq.~\eqref{eq:snapshots3} shares the same eigenvalues problem as Eq.~\eqref{eq:snapshots1}, but with lower matrix dimensions. 

Written in the vector form of $\Psi=[\bm{\psi}^1,\bm{\psi}^2,\cdots,\bm{\psi}^{N_T}]$, and rearranging $\mathbf{D}=\mathrm{diag}[d_1,d_2,\cdots, d_{N_T}]$ with order $d_1\ge d_2\ge\cdots d_{N_T}>0$, the vector form of Eq.~\eqref{eq:snapshots3} is
\begin{equation}
    \mathbf{R}\bm{\psi}^\alpha = d_\alpha\bm{\psi}^\alpha, \alpha=\{ 1, 2, \cdots,N_T\}.
    \label{eq:snapshots4}
\end{equation}

Thus, based on the limited empirical data $\mathbf{u}$, the eigenvalues and eigenfunctions are computed by the snapshots method as
\begin{equation}
\begin{cases}
    \lambda_\alpha & =d_\alpha\\
    \bm{\phi}^\alpha &\displaystyle =\frac{\bm{\psi}^\alpha}{\Vert\bm{\psi}^\alpha\Vert} = \frac{ \mathbf{Q}(\beta,\alpha)\bm{q}^\beta}{\Vert \mathbf{Q}(\beta,\alpha)\bm{q}^\beta \Vert}
\end{cases},
\end{equation}
where $\beta$ is a dummy index in $\{ 1, 2, \cdots,N_T\}$, $\mathbf{Q}(\beta,\alpha)$ is the $\beta$th row and the $\alpha$th column element of the unitary matrix $\mathbf{Q}$, and the snapshots of velocity vector $\bm{q}^\beta$ are rearranged from the velocity vector $\bm{u}^\beta$ by the Algorithm \ref{alg:a}.

The maximum cost of memory for one matrix in the snapshots method comes from the matrix $\mathbf{C}$ with $N_T\times N_T$ elements or the data vector $\bm{q}^\alpha$ with $M$ elements, both of which are far smaller than the correlation matrix $\mathbf{R}$ in Eq.~\eqref{eq:correlation} with elements $M\times M$ when $M\gg N_T$. However, the main cost of the snapshots POD method is the reading-in the data. Direct POD requires $O(N_T)$ readings of velocity data, but the snapshot method needs $O(N^2_T)$ readings of all data.


\section{Application to a statistically non-stationary lid-driven cavity simulation}\label{sec:example}

The Phase POD is here applied to a Direct Numerical Simulation (DNS) of a statistically non-stationary lid-driven cavity flow, where the two-dimensional spatial domain is defined as $\Omega_S=[0,1]\times[0,1]$, and the three-dimensional spatio-temporal domain is then $\Omega=\Omega_S\times(0,T]$. The correlation matrix is constructed according to the procedure described in Appendix \ref{appB}. The coordinate designation $(x,y)$ in the following case represents the indexed coordinates $(x_1, x_2)$, and the corresponding velocities $(u, v)$ denote the velocity components $(u_1,u_2)$.

\subsection{Direct Numerical Simulations}
\begin{figure*}[ht]
      \begin{subfigure}[b]{0.45\linewidth}
        \includegraphics[width=\textwidth]{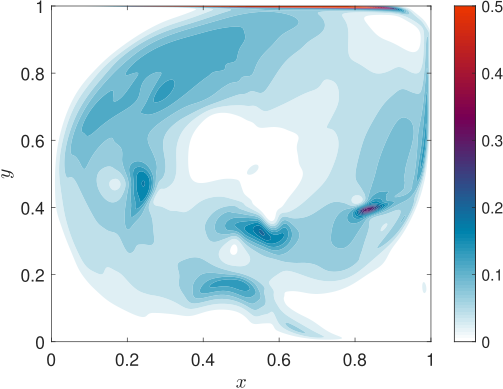}
        \caption{\label{fig:period}}
        \end{subfigure}
        \begin{subfigure}[b]{0.45\textwidth}
        \includegraphics[width=\textwidth]{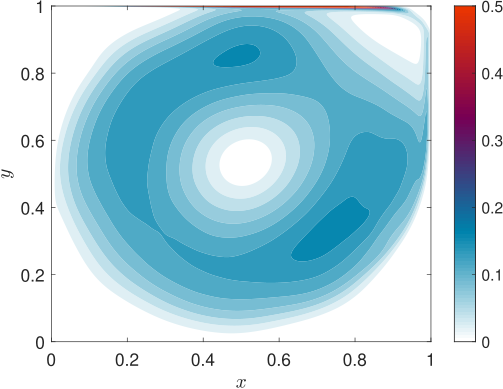}
        \caption{\label{fig:periodaverage}}
        \end{subfigure}
    \centering
    \caption{TKE contours of the statistical non-stationary lid-driven cavity at $Re = 36\,000$. (\protect\subref{fig:period}) Phase snapshot at $p=0.25T$ of the 70th period. (\protect\subref{fig:periodaverage}) Phase average at $p=0.25T$.}
\end{figure*}

The incompressible flow inside the unit-square cavity is driven by the lid velocity induced at $y=1$:
\begin{subequations}
\begin{equation}
\left.\begin{array}{rcl} 
\phantom{\dfrac{1}{1}}u(x,y=1,t)&=& 16 x^2(1-x^2)f(t)\\
v(x,y=1,t)&=& 0\end{array}\right\}\,,\quad x\in[0:1]\,,\label{eq:lid_velocity}\\
\end{equation}
\begin{equation}
f(t)=1 +\frac{1}{2} \sin(\frac{2\pi t}{T})\,,\quad t\in(0:T].\label{eq:ft}
\end{equation}
\end{subequations} 
A no-slip velocity condition is prescribed on all other walls of the cavity. For the pressure, a zero Neumann boundary condition is prescribed on all walls of the domain.
The Navier-Stokes equations are discretized using the finite-volume approach on a $512 \times 512$ Cartesian grid with non-uniform spacing, so as to resolve the boundary layers of the flow. Their solution uses the implicit pressure-velocity coupling procedure of Denner et al.\cite{Denner2020} and is second-order accurate in time and space. The fluid density is set to $1\ \mathrm{kg/m^3}$, and the viscosity is $1.67\times 10^{-4} \ \mathrm{Pa\cdot s}$. For verification purposes,  a simulation of a statistically stationary flow at $Re = 24\,000$ was first performed with $f(t)=1$ in Eq.~\eqref{eq:lid_velocity}. An ensemble of $10\,000$ snapshots were generated with a time step of $\Delta t = 0.02\mathrm{s}$. From this ensemble, the last $8\,000$ snapshots were used for a space POD analysis, and the spatial modes were found to be in good agreement with those in Balajewicz et al.\cite{Balajewicz2013} and Lee et al.\cite{Lee2019}.

The statistically non-stationary lid-driven cavity flow was then simulated using the oscillating lid velocity defined in Eq.~\eqref{eq:ft}. The period was set to $T=2\,\mathrm{s}$, and $Re = 36\,000$, defined using the amplitude of the velocity oscillation of the lid as in Vogel et al.\cite{Vogel2003}, Mendu and Das\cite{Mendu2013} and Zhu et al.\cite{Zhu2020}. A snapshot of the TKE contours for the 70th period and $p=0.25T$ is shown in Fig.~\ref{fig:period}, and the corresponding phase averaged turbulent kinetic energy at phase $p=0.25T$ is shown in Fig.~\ref{fig:periodaverage}. A total of $1\,242$ periods were simulated, each sampled into 100 phases. This corresponds to a time step $\Delta t = 0.02\mathrm{s}$, which is the same as in the steady case. The data used for the final analysis was sampled after the initial $22$ periods in order to rule out the initial transient development of the periodically forced flow. In comparison with the oscillatory lid-driven cavity flow studies by Iwatsu et al.\cite{Iwatsu1992}, Vogel et al.\cite{Vogel2003}, Blackburn and Lopez\cite{Blackburn2003} and Zhu et al.\cite{Zhu2020}, this simulation runs at a higher Reynolds number with the aim of reaching a more chaotic state.

Due to the stochastic features of turbulence, statistical convergence studies are commonly conducted prior to data processing. The uncertainties in the computed statistics of the direct numerical simulation (DNS) consist of the finite statistical sampling and the discretization of the Navier-Stokes equation. Oliver et al.\cite{Oliver2014} found that the sampling uncertainty is commonly large, while the estimated discretization errors are quite small. The simulation mesh employed in this study is more refined than in the simulations conducted by Balajewicz et al.\cite{Balajewicz2013}, Lee et al.\cite{Lee2019} and Rubini et al.\cite{Rubini2020}. We therefore neglect the discretization errors in our statistical analysis and only analyse the statistical sampling uncertainty over $1\,220$ periods of the statistically non-stationary lid-driven cavity flow in Appendix \ref{appC}. Only the velocities are considered in this sampling uncertainty analysis, because these are the only quantities used in the Phase POD to compose the correlation matrix in this statistically non-stationary lid-driven cavity simulation. The maximum variability of both velocity components was found to be below $2\%$. The Reynolds stress tensor was evaluated along the diagonal of the square domain over all phases, and was shown to converge sufficiently with increasing $n$ over all phases. Moffat\cite{Moffat1988} and Coleman and Steele\cite{Coleman1995} stated that a 95\% confidence interval can provide understandable and reproducible experimental statistics. The data set produced by our DNS can thus be considered acceptable for analysis using the Phase POD method.

\subsection{Modal dynamics}

\begin{figure*}[ht]
        \begin{subfigure}[b]{0.28\textwidth}
        \includegraphics[height=4cm, width=4cm]{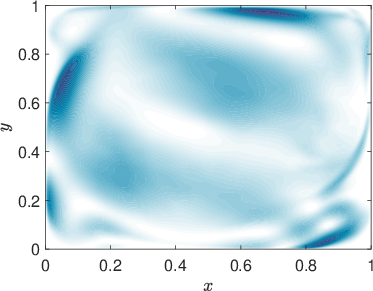}
        \caption{\label{fig:mode1p30}}
        \end{subfigure}
        \hfill
        \begin{subfigure}[b]{0.28\textwidth}
        \includegraphics[height=4cm, width=4cm]{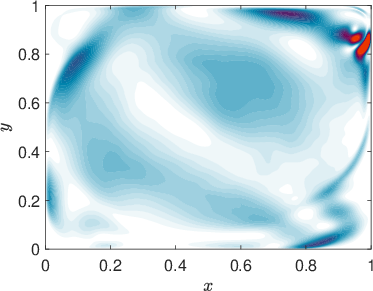}
        \caption{\label{fig:mode1p60}}
        \end{subfigure}
        \hfill
        \begin{subfigure}[b]{0.298\textwidth}
        \includegraphics[height=4.2cm, width=5cm]{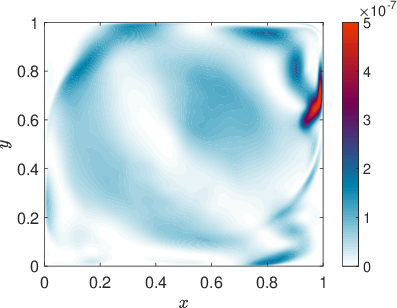}
        \caption{\label{fig:mode1p90}}
        \end{subfigure}
        \begin{subfigure}[b]{0.28\textwidth}
        \includegraphics[height=4cm, width=4cm]{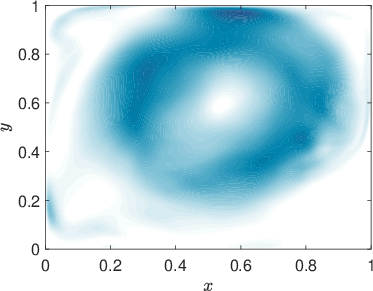}
        \caption{\label{fig:mode2p30}}
        \end{subfigure}
        \hfill
        \begin{subfigure}[b]{0.28\textwidth}
        \includegraphics[height=4cm, width=4cm]{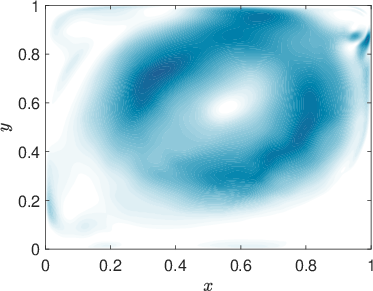}
        \caption{\label{fig:mode2p60}}
        \end{subfigure}
        \hfill
        \begin{subfigure}[b]{0.298\textwidth}
        \includegraphics[height=4.2cm, width=5cm]{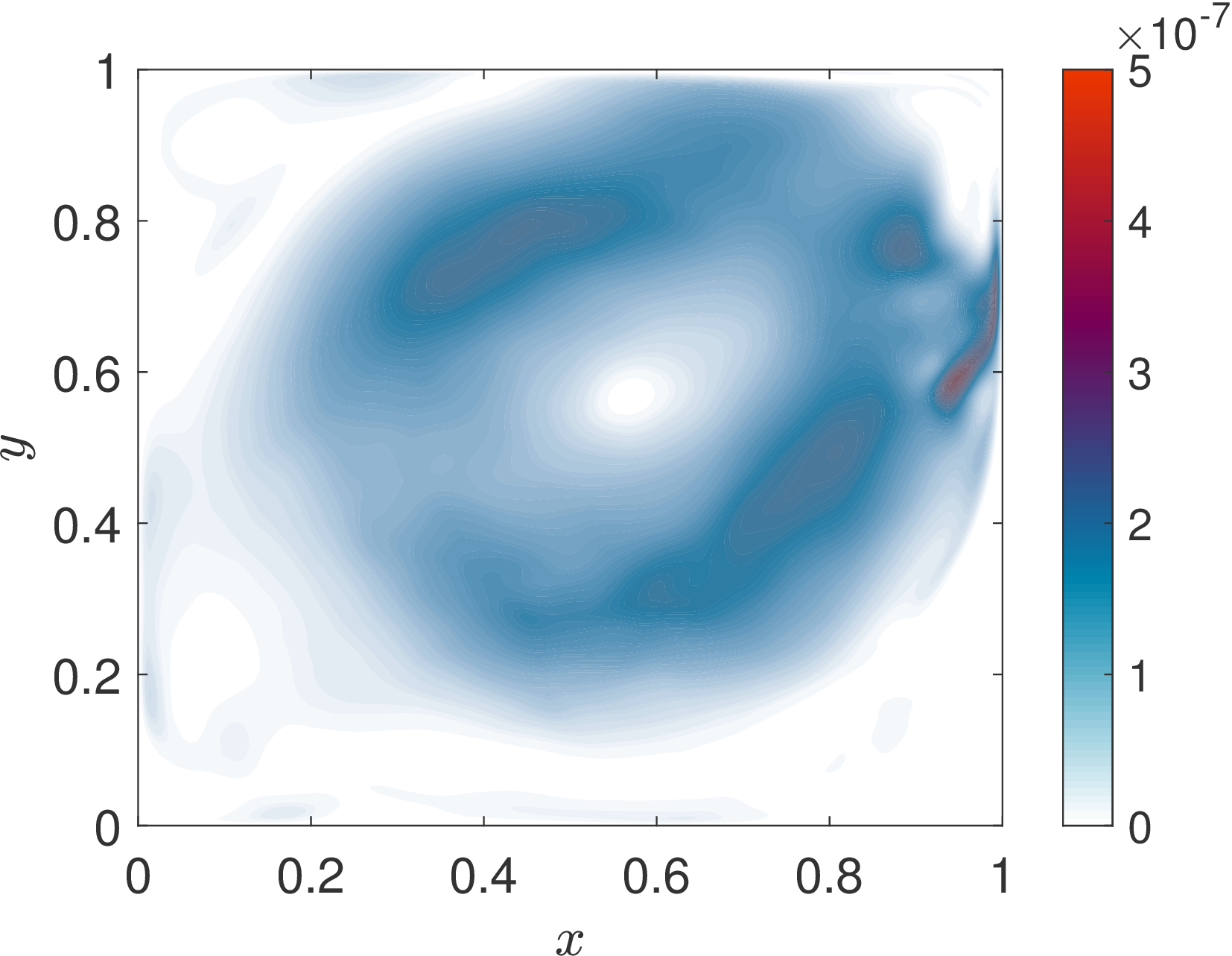}
        \caption{\label{fig:mode2p90}}
        \end{subfigure}
        \begin{subfigure}[b]{0.28\textwidth}
        \includegraphics[height=4cm, width=4cm]{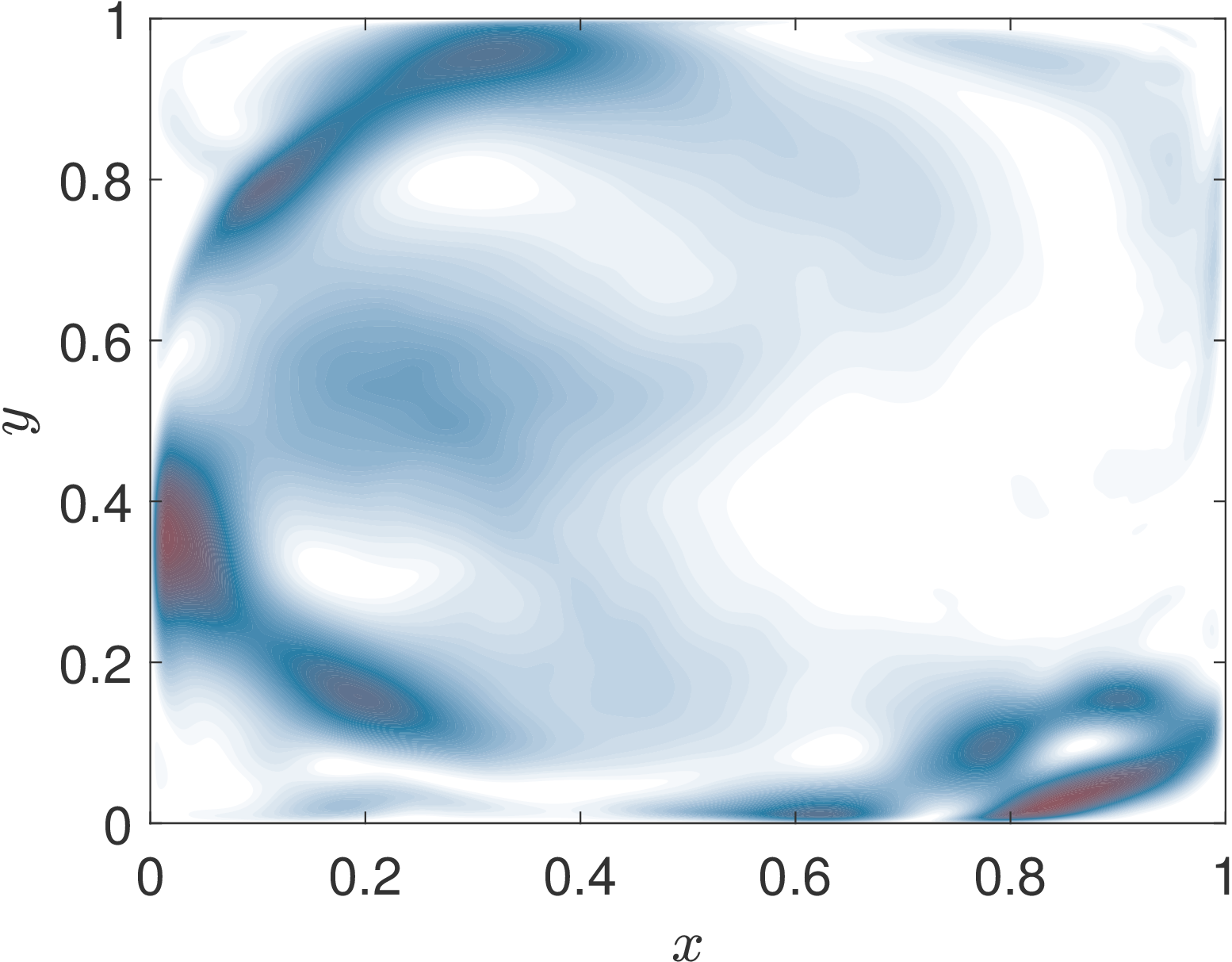}
        \caption{\label{fig:mode3p30}}
        \end{subfigure}
        \hfill
        \begin{subfigure}[b]{0.28\textwidth}
        \includegraphics[height=4cm, width=4cm]{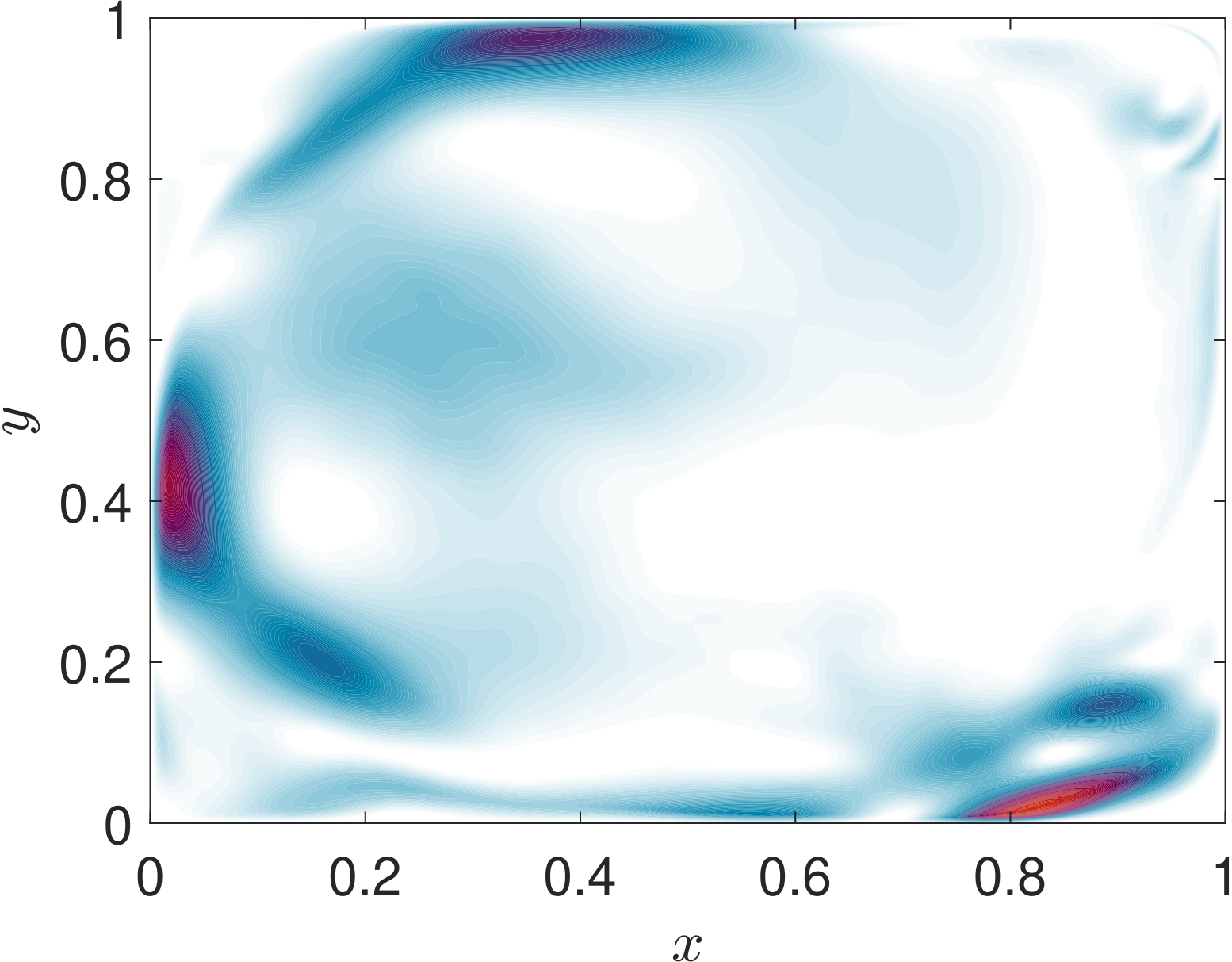}
        \caption{\label{fig:mode3p60}}
        \end{subfigure}
        \hfill
        \begin{subfigure}[b]{0.298\textwidth}
        \includegraphics[height=4.2cm, width=5cm]{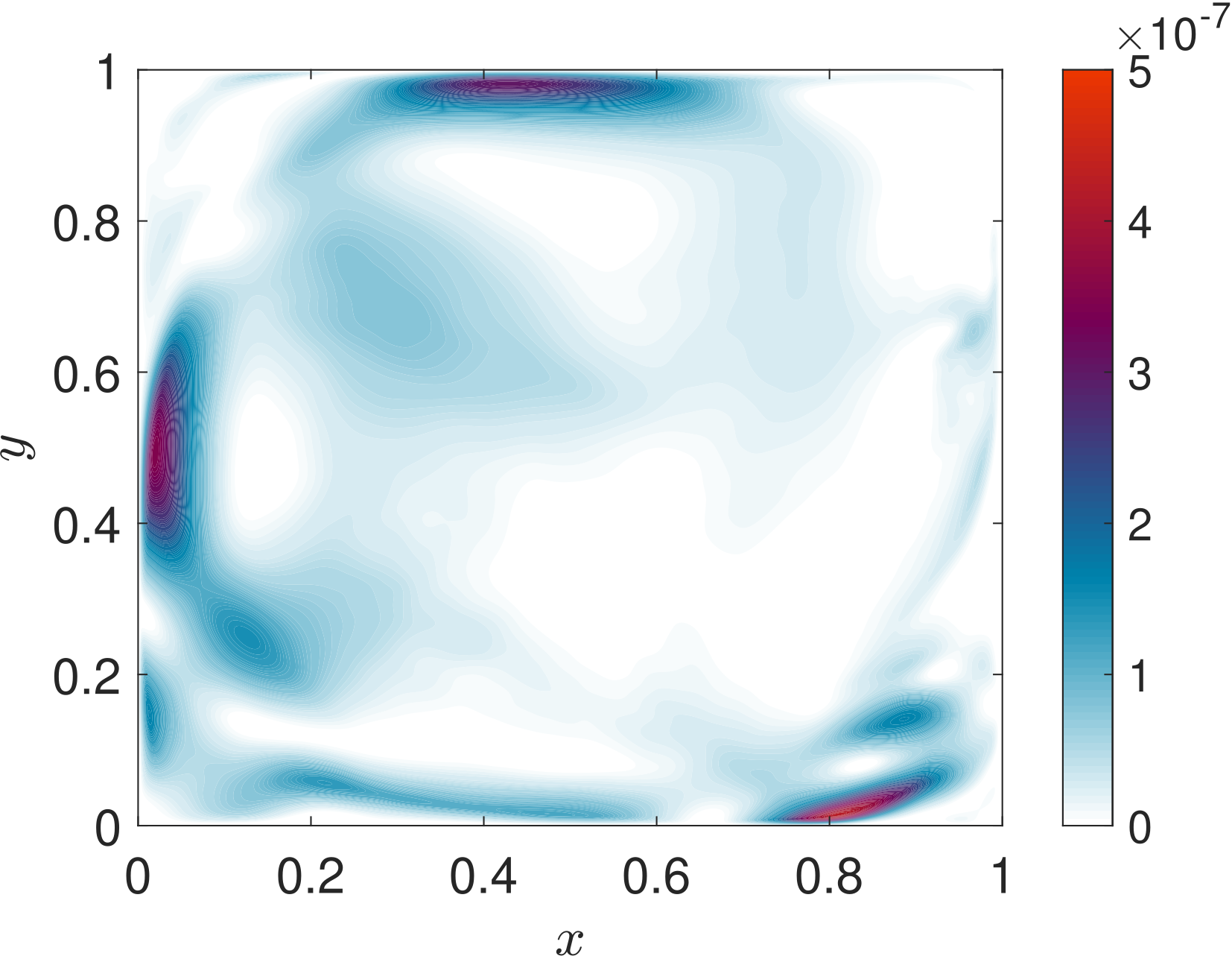}
        \caption{\label{fig:mode3p90}}
        \end{subfigure}
        \begin{subfigure}[b]{0.28\textwidth}
        \includegraphics[height=4cm, width=4cm]{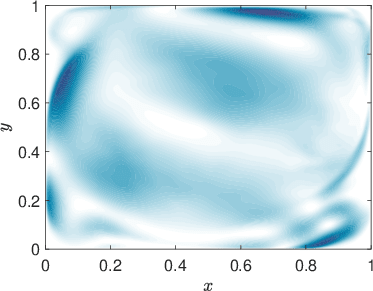}
        \caption{\label{fig:mode4p30}}
        \end{subfigure}
        \hfill
        \begin{subfigure}[b]{0.28\textwidth}
        \includegraphics[height=4cm, width=4cm]{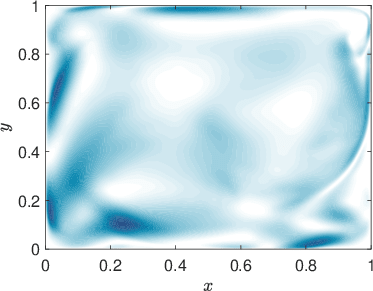}
        \caption{\label{fig:mode4p60}}
        \end{subfigure}
        \hfill
        \begin{subfigure}[b]{0.298\textwidth}
        \includegraphics[height=4.2cm, width=5cm]{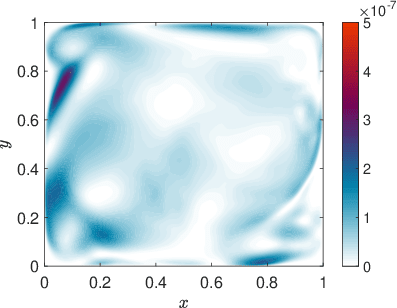}
        \caption{\label{fig:mode4p90}}
        \end{subfigure}
    \centering
    \caption{Iso-contours of the oscillating lid driven cavity Phase POD TKE modes 1-4, $\varphi^\alpha_1\varphi^{1*}_\alpha+\varphi^\alpha_2\varphi^{2*}_\alpha$, with $\alpha\in \{1,2,3,4\}$. Mode 1: (\protect\subref{fig:mode1p30}) $0.3T$, (\protect\subref{fig:mode1p60}) $0.6T$, (\protect\subref{fig:mode1p90})$0.9T$; mode 2: (\protect\subref{fig:mode2p30}) $0.3T$, (\protect\subref{fig:mode2p60}) $0.6T$, (\protect\subref{fig:mode2p90})$0.9T$; mode 3: (\protect\subref{fig:mode3p30}) $0.3T$, (\protect\subref{fig:mode3p60}) $0.6T$, (\protect\subref{fig:mode3p90})$0.9T$; mode 4: (\protect\subref{fig:mode4p30}) $0.3T$, (\protect\subref{fig:mode4p60}) $0.6T$, (\protect\subref{fig:mode4p90})$0.9T$. The videos view by: https://doi.org/10.11583/DTU.21444684.\cite{Zhang2022}}
    \label{fig:modesplot}
\end{figure*}
Then boundary condition given in Eq.~\eqref{eq:ft} injects energy into the cavity flow through two main contributions: (1) the steady part of $f(t)$, equals to 1, that sustains the basic turbulent flow; (2) the non-stationary part of $f(t)$, equals to $\frac{1}{2}\sin(\frac{2\pi t}{T})$, that introduces time variations by periodic modulation. Phase POD is designed to extract the dynamics generated by both contributions. 

Iso-contours of the oscillating lid-driven cavity TKE Phase POD modes 1-4 for the three phases $\{0.3T, 0.6T, 0.9T\}$ are presented in Fig.~\ref{fig:modesplot} to provide insight into the turbulence structure in the cavity. The red contours represent regions of the plane where the TKE is greater than the local mean TKE and regions of blue colour indicate a TKE smaller than the local mean. In general, from these four modes one can observe two main processes exchanging energy; (1) an inner flow rotation that exchanges energy through shear with (2) an outer flow rotation with significant shear due to the presence of both the bounding walls and the inner flow rotational motion.

A movie showing the spatio-temporal behavior of modes 1-4, corresponding to Fig.~\ref{fig:modesplot}, is available at \url{https://doi.org/10.11583/DTU.21444684}\cite{Zhang2022} and on the web page of the DTU Turbulence Research Lab. Under the same link, a movie of the full dynamics of modes \{1,2,3,4,5,6,7,10,20,50,70,100,150,200\} is also available for the readers' reference.

Fig.~\ref{fig:mode1p30}-\ref{fig:mode1p90} illustrate how mode 1 varies with the non-stationary boundary conditions (the lid velocity modulation) across the three phases $p=\{0.3T,0.6T,0.9T\}$ (left to right) presented. The time-varying energy injected into the cavity from the top lid is observed to accumulate near the top-right corner, which results in separation towards the end of the phase period. One region of significant energy variation across phase is apparent in the central region, $(x,y) \approx (0.6,0.7)$, where mode 2 also has a local minimum. The TKE variations with phase in the high shear regions close to the walls is also significant. 

In Fig.~\ref{fig:mode2p30}-\ref{fig:mode2p90}, the energetic inner structure of mode 2 is observed to be rotating in the central region while exchanging energy with structures from the corners, its rotational center and the structures along the boundaries (in particular the right boundary). Further evidence of interactions between the inner region and the structures near the boundaries can be found in mode 3 and mode 4. Fig.~\ref{fig:mode3p30}-\ref{fig:mode3p90} illustrates that mode 3 exchanges energy with the inner rotating structure, in particular near the left and top boundaries, and the energy in the right-bottom corner separates into two branches. Mode 4 displays energetic structures closer to the walls and with more clear energy separation in the corners in Fig.~\ref{fig:mode4p30}-\ref{fig:mode4p90}. 

Summarizing the energy transport behavior of the different modes in Fig.~\ref{fig:modesplot} and the movies in the supplementary data, four significant patterns are found to form and transfer turbulent structures from the view of turbulent kinetic energy in the modes. These patterns illustrate the dynamics of the maximum TKE as a function of phase. Here, we only discuss modes \{1,2,3,4\} to illustrate how one may analyse the dynamics of the Phase POD modes. 

\begin{figure*}[ht]
        \begin{subfigure}[b]{0.48\textwidth}
        \includegraphics[width=\textwidth]{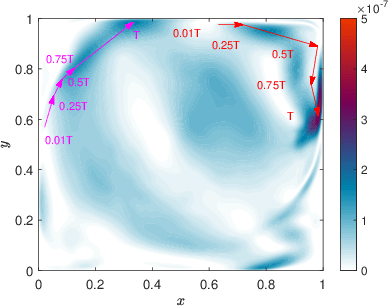}
        \caption{\label{fig:pattern1}}
        \end{subfigure}
        \begin{subfigure}[b]{0.48\textwidth}
        \includegraphics[width=\textwidth]{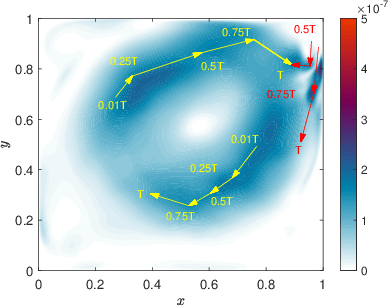}
        \caption{\label{fig:pattern2}}
        \end{subfigure}
        \begin{subfigure}[b]{0.48\textwidth}
        \includegraphics[width=\textwidth]{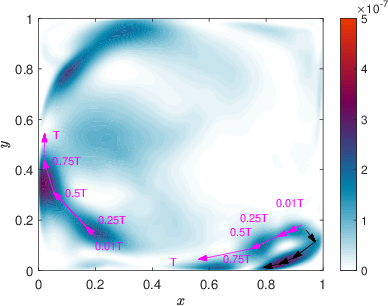}
        \caption{\label{fig:pattern3}}
        \end{subfigure}
        \begin{subfigure}[b]{0.48\textwidth}
        \includegraphics[width=\textwidth]{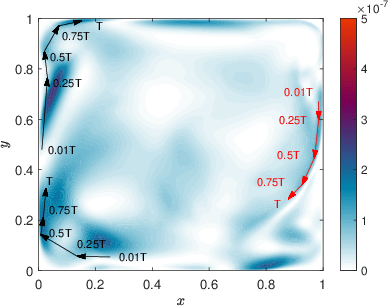}
        \caption{\label{fig:pattern4}}
        \end{subfigure}
    \caption{The four patterns of energy dynamics in the statistically non-stationary lid-driven cavity, $\varphi^\alpha_1\varphi^{1*}_\alpha+\varphi^\alpha_2\varphi^{2*}_\alpha$. Pattern 1 -- red arrows, pattern 2 -- yellow arrows, pattern 3 -- magenta arrows and pattern 4 -- black arrows. (\protect\subref{fig:pattern1}) The TKE contours of mode 1 for $p=T$. (\protect\subref{fig:pattern2}) The TKE contours of mode 2 for $p=0.75T$. (\protect\subref{fig:pattern3}) The TKE contours of mode 3 for $p=0.75T$. (\protect\subref{fig:pattern4}) The TKE contours of mode 4 for $p=0.75T$.}
    \label{fig:patterns}
\end{figure*}

\textit{Pattern 1: impact of the non-stationary part (red arrows in Fig.~\ref{fig:patterns})}. Pattern 1 illustrates the impact of the non-stationary lid velocity on both the inner rotating region and on the flow near the right bounding wall. The red arrows and red text in Fig.~\ref{fig:pattern1}, Fig.~\ref{fig:pattern2} and Fig.~\ref{fig:pattern4} show how Pattern 1 develops. Firstly, as shown in Fig.~\ref{fig:pattern1}, the TKE injected by the oscillatory part of the motion of the top lid results in an increase of TKE in the upper right corner that continues to spread down along the right boundary of the cavity. Secondly, as shown in mode 2 in Fig.~\ref{fig:pattern2}, the high TKE region in the upper right corner bifurcates into two sub-structures. These two sub-structures interact with the inner rotating structure (pattern 2 described in the next paragraph). Finally, in mode 4 in Fig.~\ref{fig:pattern4}, after impacting with the right bounding wall, the remaining TKE of pattern 1 is observed to partially merge with the inner rotating structure and partially contribute to the rotational separation motion in the lower right corner.

\textit{Pattern 2: Stable center rotation (yellow arrows in Fig.~\ref{fig:patterns})}. Pattern 2 in Fig.~\ref{fig:pattern2} illustrates the relatively stable high energy rotating structures in the center region. It can be seen that the clockwise rotational motion of the high TKE structure in the inner region is reminiscent of the average TKE in Fig.~\ref{fig:periodaverage}. Since all other identified patterns (1, 3 and 4) interact with this central inner rotational structure, pattern 2 in fact consists of a band of structures with weak connection. Due to the modulation added on top of the uniform lid velocity, the rotational speed of the inner rotational structure is also not uniform and follows the dynamics of the lid velocity.

\textit{Pattern 3: Interaction with bottom and left boundaries (magenta arrows in Fig.~\ref{fig:patterns})}. Pattern 3 in Fig.~\ref{fig:pattern1} and Fig.~\ref{fig:pattern3} illustrates the interactions between the structures in the high shear region, between the inner rotating structure near the bottom and left boundaries. These modal dynamics are less tightly coupled to the non-stationary part of $f(t)$. In particular, during the beginning of the phase cycle, part of the TKE in the central region (pattern 2) introduces low TKE structures near the bottom and left boundaries. With increasing phase, low TKE structures accumulate near the bottom and left boundaries to form higher TKE structures. With further increase in phase, the accumulated TKE structures leave these boundaries, deform and transform into lower TKE structures, moving toward the top boundary in higher order modes. This process also introduces separation, in this case in the two lower corners (characterised by pattern 4).

\textit{Pattern 4: Separation in singular corner (black arrows in Fig.~\ref{fig:patterns})}. Separation occurs as the thin wall-parallel flow structures shift past one wall boundary to the next. Pattern 4 illustrates how smaller elongated structures, separated from the large inner structure rotating near the center (pattern 2), travel along the walls towards the corners where separation occurs. The TKE of the elongated wall-parallel separated structures are relatively low at the beginning of the phase, but grow as they approach the next boundary.

In the commonly applied space POD analysis (in the special case of stationary cavity flow), it is not possible to analyse the dynamical generation and interactions of these structures. The Phase POD provides such a dynamic view of the coherent structures. Patterns 3 and 4 can explain how the modes in steady lid-driven cavity flow are formed near the boundaries and the singular corners. Higher-order modes of the non-stationary cavity flow repeat these four patterns in the corresponding region with higher-order structures.

Higher-order modes have higher-order structures, which are higher frequency in time or smaller structures in space. To show the higher-order structures, a wavenumber or a frequency is need to define a spatiotemporal Phase POD mode. In practice, based on the mean frequency definition in the Hilbert Spectral Analysis (HSA) \cite{Huang1998} and the Empirical Modes Decomposition (EMD) method\cite{Flandrin2004}, there are three ways to define the mean frequency of each mode. The first one is proposed by Huang \cite{Huang1998}, which use the Fourier spectrum to compute the mean frequency; the second one is introduced by Flandrin and Gonçalvès\cite{Flandrin2004}, who take the zero-crossing number and the distance between the first and last zero-crossing to obtain the mean frequency; the third one is computed from the Hilbert marginal spectrum by Huang \cite{Huang2009}. The two energy-weighted estimators (the first one and the last one) give almost the same mean frequency\cite{Huang2009}.

\begin{figure*}[ht]
    \begin{subfigure}[b]{0.46\textwidth}
        \includegraphics[width=\textwidth]{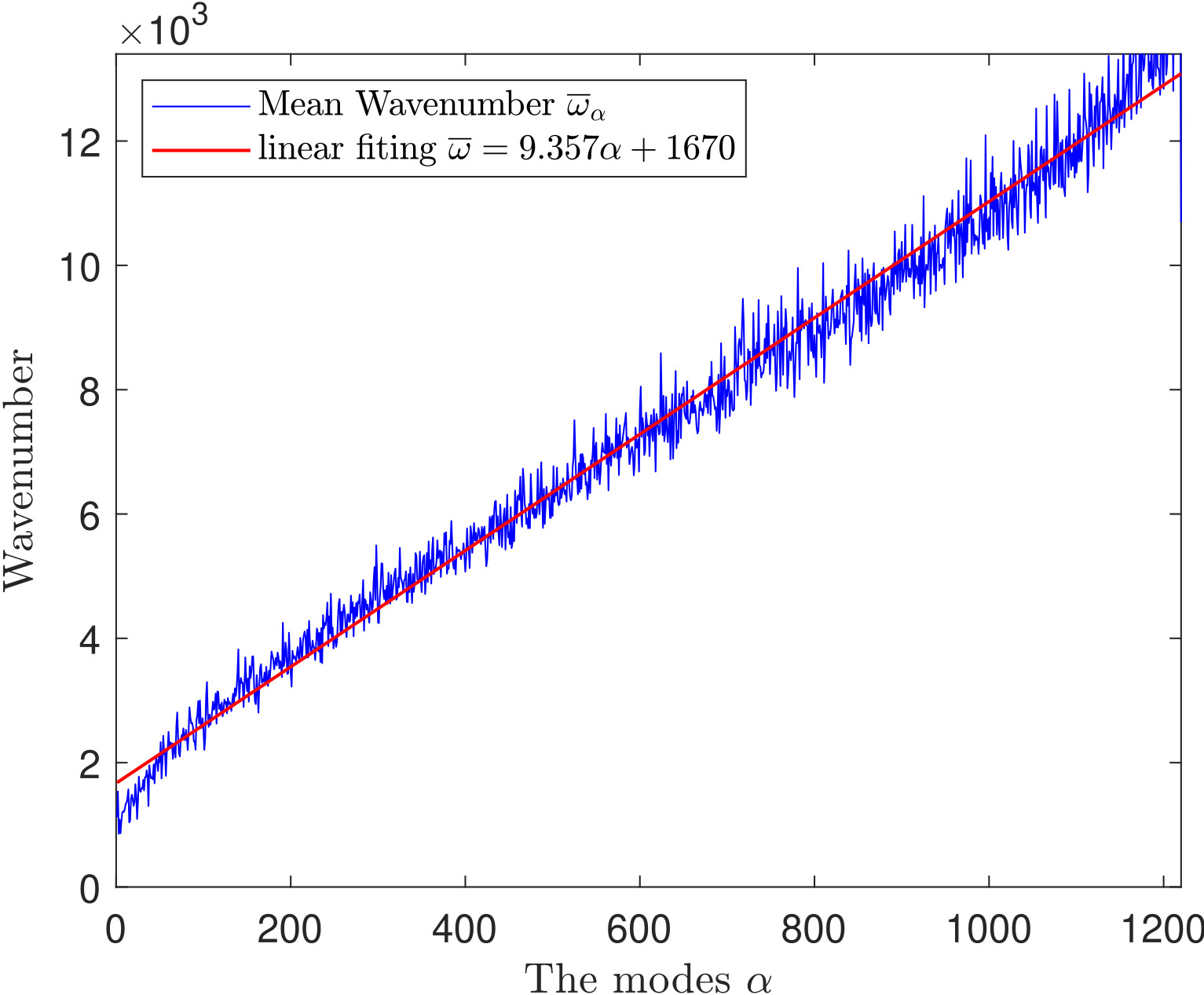}
        \caption{\label{fig:meanwavenumbermode}}
    \end{subfigure}
    \hfill
    \begin{subfigure}[b]{0.46\textwidth}
        \includegraphics[width=\textwidth]{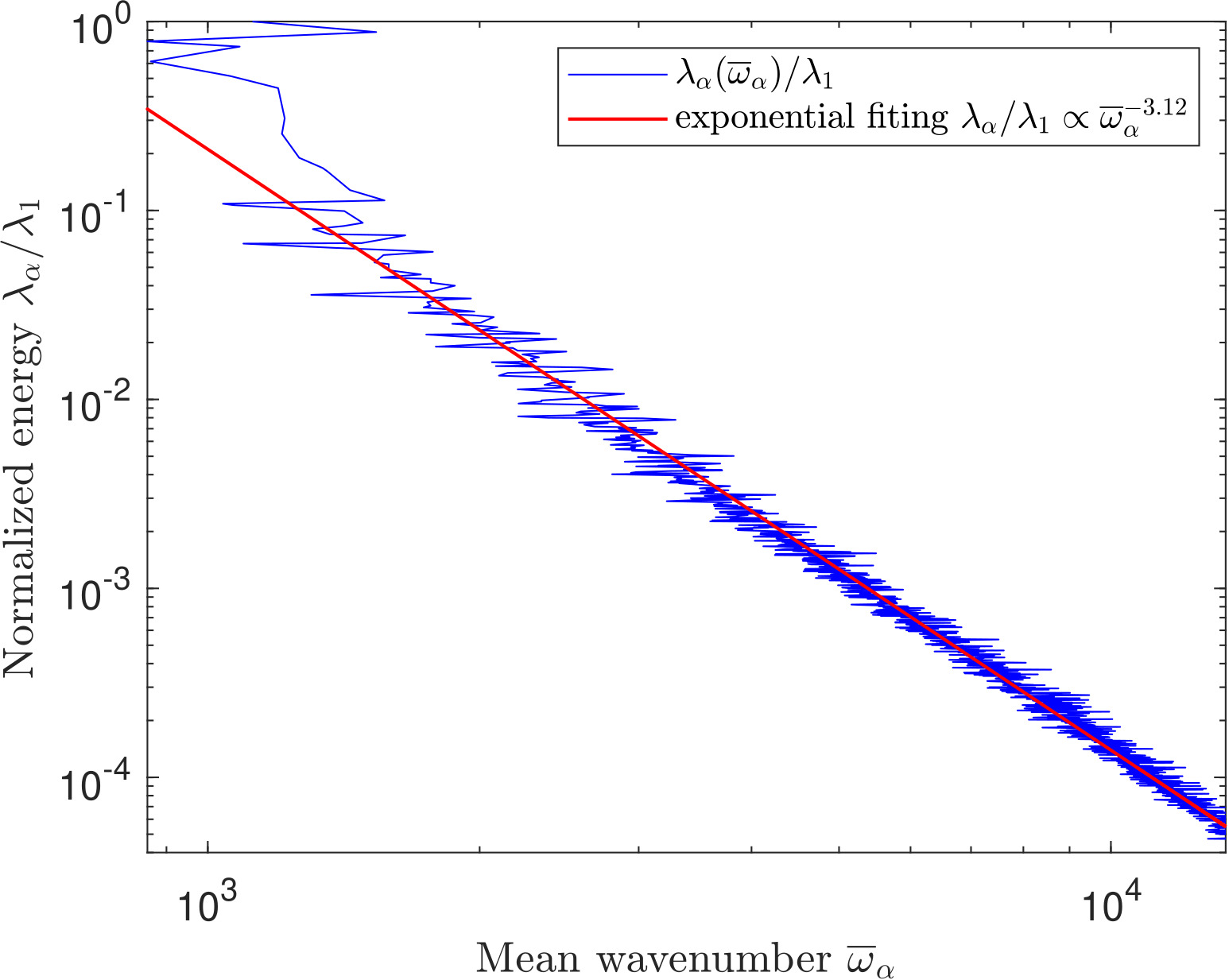}
        \caption{\label{fig:meanwavenumberenergy}}
    \end{subfigure}

    \caption{The mean wavenumber. (\protect\subref{fig:meanwavenumbermode}) Representation of the mean frequency $\overline{\omega_\alpha}$ vs. the mode index $n$ in a linear-linear representation. (\protect\subref{fig:meanwavenumberenergy}) Representation of energy spectrum with the normalized energy $\lambda_{\alpha}/\lambda_1$ vs. the mean frequency $\overline{\omega_\alpha}$ in a log-log representation.}
    \label{fig:meanwavenumber}
\end{figure*}

The definition of the mean frequency in HSA and EMD is based on the temporal modes, however the Phase POD modes include both temporal dimension and spatial dimensions. We introduce a new definition of the mean wavenumber from the HSA definition by Huang \cite{Huang1998} as 
\begin{equation*}
    \overline{\omega}_\alpha = \frac{\int_{\Omega} \omega_1 \omega_2 \omega_3 f_t S_\alpha(\bm{\omega}) d\omega_1 d\omega_2 d\omega_3 df_t }{\int_{\Omega}S_\alpha(\bm{\omega}) d\omega_1 d\omega_2 d\omega_3 df_t},
\end{equation*}
where $\bm{\omega}=\{\omega_1, \omega_2, \omega_3, f_t \}$ is a wavenumber vector with three spatial wavenumbers $\omega_1$, $\omega_2$, $\omega_3$ and one temporal frequency $f_t$. $S_\alpha(\bm{\omega})$ is the four-dimensional Fourier spectrum of the Phase POD mode $\varphi^\alpha$. Each Phase POD mode consists of several dominant wavenumbers in both the spatial coordinates and the temporal coordinate. It is impossible to pick one wavenumber in all the wavenumbers to represent the Phase POD mode. The mean wavenumber gives an energy-weighted estimation of the Phase POD modes wavenumber. Graphically, the mean wavenumber suggests a linear fitting law with the mode number $\alpha$ in Fig.~\ref{fig:meanwavenumbermode}. The higher-order Phase POD modes correspond to higher wavenumbers. Note that the higher-order Phase POD modes represent more chaotic structures, and are not the same as the HSA modes and EMD modes, where higher-order Intrinsic Mode Function (IMF) modes show more smooth behaviours. So it is reasonable that the mean wavenumber shows a linear trend with the mode index $\alpha$.

The energy spectrum shows an exponential trend with scaling exponent -3.12 in Fig.~\ref{fig:meanwavenumberenergy}, where $\lambda_\alpha$ is the eigenvalue of the mode $\varphi_\alpha$ representing the TKE of the mode $\alpha$. The scaling exponent -3.12 in this two-dimensional non-stationary lid-driven flow is steeper than the inertial range scaling power -5/3 in the stationary turbulent flows.

\subsection{Spectral energy budget}
In the studies by Muralidhar et al.\cite{Muralidhar2019} and Couplet et al.\cite{Couplet2003}, the contribution of the dominant modes were examined through their contribution to the energy transport equation. Muralidhar et al.\cite{Muralidhar2019} show turbulent transport terms with the quadratic interactions in Fourier space, and the goal of the current study is to demonstrate the effects of non-stationarity on the spectral energy transport equation.

The material derivative of the TKE expresses the TKE dynamics. Any velocity vector form of Eq.~\eqref{eq:u} is re-written as a combination of the basis,
\begin{equation} \label{eq:udecom}
\bm{q}^\beta=\left[
\begin{array}{c}
  \bm{q}_1^\beta\\[1ex]
  \bm{q}_2^\beta\\[1ex]
  \bm{q}_3^\beta
\end{array}
\right] 
=\sum^\infty_\alpha a_\alpha \varphi^\alpha=\sum^\infty_\alpha a_\alpha 
\left[
\begin{array}{c}
 \varphi_1^\alpha\\[1ex]
 \varphi_2^\alpha\\[1ex]
 \varphi_3^\alpha
\end{array}
\right].
\end{equation}
where $\varphi_1$, $\varphi_2$ and $\varphi_3$ denote mode components in the coordinate $(x_1, x_2, x_3)$. Using $E$ as the TKE, $E=\frac{1}{2}(u_1^2+u_2^2+u_3^2)$, the equation of TKE for incompressible flow can be obtained directly from the Reynolds-stress equation by contracting the free indices \cite{George2013, Azur2018}:
\begin{equation}\label{eq:TKEequation}
\begin{split}
    \underbrace{\frac{\partial E}{\partial t}}_{\text{I.a}} + \underbrace{\langle \tilde{u}_j\rangle \nabla_j E}_{\text{I.b}} + \underbrace{\langle u_i u_j\rangle \nabla_j  \langle \tilde{u}_i\rangle}_{\text{II}} = -\underbrace{\frac{1}{2}\nabla_j \langle  u_i u_i u_j\rangle}_{\text{III}}\\
    -\underbrace{\frac{1}{\rho} \nabla_i \langle p u_i\rangle}_{\text{IV}} +\underbrace{2\nu\nabla_j \langle  s_{ij} u_i\rangle}_{\text{V}} - \underbrace{2\nu\langle s_{ij} s_{ij}\rangle}_{\text{VI}}\\
\end{split}
\end{equation}
where $s_{ij}=\frac{1}{2}(\nabla_j u_i+\nabla_i u_j)$ is the strain tensor of fluctuating velocity.

Term I.a is the rate of change of TKE due to non-stationarity. Term I.b is the rate of change of TKE due to advection by the mean flow through an inhomogeneous field. Term II represents the turbulence-energy production by mean shear. Term III is the transport term due to velocity fluctuations. Term IV is the transport term due to pressure fluctuations. Term V is the viscous transport and term VI is the viscous dissipation. Terms I.a and I.b constitute the material derivative of the TKE, and terms from III to V represent the transport of TKE. The viscosity only influences Terms V and VI.

Expanding the energy equation in terms of the basis functions Eq.~\eqref{eq:udecom} yields the following expressions for the terms (see p. 45 in Ref.~\onlinecite{Azur2018}):
\begin{equation}\label{eq:TKEmodes}
    \begin{cases}
    \text{I.a: } \displaystyle\frac{\lambda_\alpha}{2}\frac{\partial}{\partial t} (\varphi^\alpha_i\varphi^{\alpha*}_i)\\[2ex]
    \text{I.b: } \lambda_\alpha\langle \tilde{u}_j\rangle \nabla_j ( \varphi^\alpha_i\varphi^{\alpha*}_i)\\[2ex]
    \text{II: } \lambda_\alpha\varphi^{\alpha}_i\varphi^{j*}_\alpha\nabla_j\langle \tilde{u}^i\rangle\\[2ex]
    \text{III: } \displaystyle -\frac{1}{2}\langle a_\alpha a^{\beta*} a_\gamma\rangle\varphi^j_\gamma \nabla_j ( \varphi^\alpha_i\varphi^{i*}_\beta)\\[2ex]
    \text{IV: } \displaystyle \frac{1}{\rho} \varphi^{\alpha*}_i\langle a_\alpha\nabla_i p\rangle\\[2ex]
    \text{V: } \nu\lambda_\alpha \nabla_j (\varphi^\alpha_i\nabla^j\varphi^{i*}_\alpha +\varphi^{\alpha}_i\nabla^i\varphi^{j*}_\alpha)\\[2ex]
    \text{VI: }  \nu\lambda_\alpha(\nabla^j \varphi^{i\alpha}\nabla_j\varphi^{*}_{i\alpha} +\nabla^i\varphi^{j\alpha}\nabla_j\varphi^{*}_{i\alpha})
    \end{cases}.
\end{equation}

In the 2D statistically non-stationary lid-driven cavity flow, the velocity vector is $\bm{q}^\beta= [\bm{q}_1^\beta, \bm{q}_2^\beta]^T =\sum^\infty_\alpha a_\alpha [ \varphi_1^\alpha,\varphi_2^\alpha]^T$, and the TKE is $E=\frac{1}{2}(u^2_1+u^2_2)$. The free indices $i$ and $j$ in Eq.~\eqref{eq:TKEequation} and Eq.~\eqref{eq:TKEmodes} are in $\{1,2\}$.

\subsection{Triadic interactions between modes}
\begin{figure*}
    \centering
    \begin{subfigure}[b]{\textwidth}
        \includegraphics[width=0.46\textwidth]{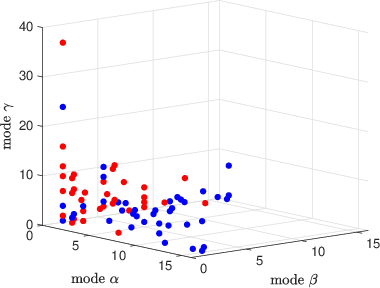}
        \caption{\label{fig:triad1}}
    \end{subfigure}
    \begin{subfigure}[b]{0.46\textwidth}
        \includegraphics[width=\textwidth]{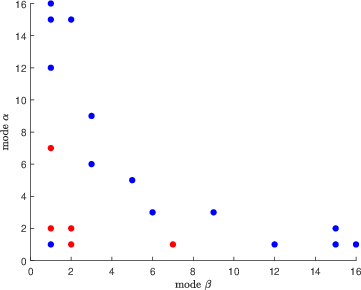}
        \caption{\label{fig:triad11}}
    \end{subfigure}
    \hfill
    \begin{subfigure}[b]{0.46\textwidth}
        \includegraphics[width=\textwidth]{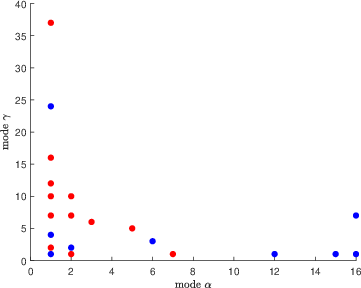}
        \caption{\label{fig:triad12}}
    \end{subfigure}
        
     \caption{Distribution of the 34 most energetic terms (absolute value) computed by $\int_{\Omega} \mathcal{T}(\alpha, \beta, \gamma) d\bm{x}$. Blue dots correspond to negative energy transfer (and negative dissipation for the corresponding combination $(\alpha, \beta, \gamma)$); Red dots correspond to positive energy transfer, meaning that energy increases for the corresponding combination of modes. (\protect\subref{fig:triad1}) 3D scattering distribution of the 34 most energetic terms. (\protect\subref{fig:triad11}) A slice of sub-figure (\protect\subref{fig:triad1}) where $\gamma=1$. (\protect\subref{fig:triad12}) A slice of sub-figure (\protect\subref{fig:triad1}) where $\beta=1$.}
    \label{fig:triad0}
\end{figure*}
Analogously to the quadratic term of triple interactions in the study of a pipe flow by Muralidhar et al.\cite{Muralidhar2019}, computing and analyzing the correlations of Term III is problematic because it not only requires the triple correlations of the coefficients $\langle a_\alpha a^{\beta*} a_\gamma\rangle$, but also involves nonlinear triads of modes with first derivatives $\varphi^j_\gamma \nabla_j ( \varphi^\alpha_i\varphi^{i*}_\beta)$. However, we can simplify this term by defining an interaction term $\bm{k}^\alpha_\beta$ between mode $\alpha$ and mode $\beta$ as $\bm{k}^\alpha_\beta=\varphi^\alpha_1\varphi^{1*}_\beta+\varphi^\alpha_2\varphi^{2*}_\beta+\varphi^\alpha_3\varphi^{3*}_\beta$. Term III (defined as $\mathcal{T}(\alpha, \beta, \gamma)$ in the following text) summed over all velocity components with $i$ is
\begin{eqnarray*}
\mathcal{T}(\alpha, \beta, \gamma) &=& -\frac{1}{2}\langle a_\alpha a^{\beta*} a_\gamma\rangle\varphi^j_\gamma \nabla_j ( \varphi^\alpha_i\varphi^{i*}_\beta)\\
&=&-\frac{1}{2}\langle a_\alpha a^{\beta*} a_\gamma\rangle\varphi^j_\gamma \nabla_j \bm{k}^\alpha_\beta.
\end{eqnarray*}
In this way, $\bm{k}^{\alpha}_\alpha$ denotes the energy structure related to mode $\alpha$, and $\varphi^j_\gamma \nabla_j \bm{k}^\alpha_\alpha$ is then the convection of this energy structure. The coefficient term $\langle a_\alpha a^{\beta*} a_\gamma\rangle=\langle (a_\alpha a^{\alpha*}) a_\gamma\rangle$ means instant turbulent energy of mode $\alpha$ projected on the instant coefficient of mode $\gamma$.

Similar to the interpretation by McKeon\cite{McKeon2017} that the nonlinearity provides the forcing to the linear Navier–Stokes operator under the resolvent framework, the turbulence transport term III means that nonlinear triadic interactions provide the forcing to the energy of each mode that is running independently under the dynamic system governed by the Navier-Stokes equation. The mode $\gamma$ produces a convection effect on the interaction between mode $\alpha$ and mode $\beta$. The absolute value of the $34$ most energetic modal combination terms $(\alpha, \beta, \gamma)$ are shown in Fig.~\ref{fig:triad0}. It can be seen that in the lower $\gamma$ modes, higher interaction modes (higher than 5) of $(\alpha,\beta)$ tend to be negative, and in the higher $\gamma$ modes (higher than 5), the lower interaction modes (lower than 5) of $(\alpha,\beta)$ tend to be positive. The energetic exchanges between the modes via triadic interaction consist of two parts and are discussed in further detail below: (1) interaction (or projection) between modes $\alpha$ and $\beta$ after convection of all modes; (2) convection effect of mode $\gamma$ on the interactions of mode $\alpha$ and $\beta$.

\begin{figure*}[ht]
    \begin{subfigure}[b]{0.46\textwidth}
        \includegraphics[width=\textwidth]{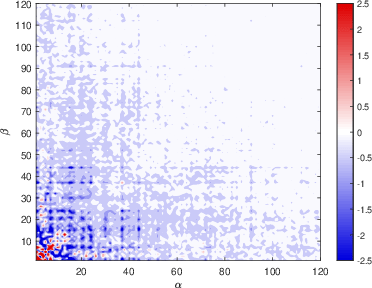}
        \caption{\label{fig:triad3}}
    \end{subfigure}
    \hfill
    \begin{subfigure}[b]{0.46\textwidth}
        \includegraphics[width=\textwidth]{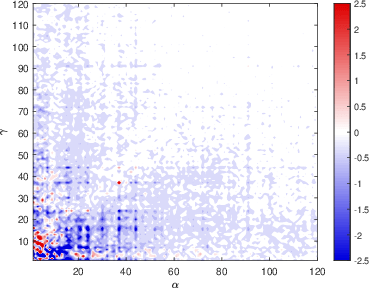}
        \caption{\label{fig:triad2}}
    \end{subfigure}

    \caption{Distribution of (\protect\subref{fig:triad3}) $\Xi(\alpha, \beta)$ and (\protect\subref{fig:triad2}) $\Pi(\gamma\vert\alpha)$.  The color coding is concurrent with that of Fig.~\ref{fig:triad0}.}
    \label{fig:pixi}
\end{figure*}

(1) {\it Interactions between modes $\alpha$ and $\beta$ after convection of all modes.} Each mode can produce a convection effect on the interaction of any pair of modes. However, the high energy modes may generate stronger convection effects due to the higher absolute values of their coefficients. The influence of all convective modes on the energy interaction mode pair $(\alpha, \beta)$ is given by
\begin{equation}\label{eq:xi}
    \Xi(\alpha, \beta)=\sum_{\gamma} \int_{\Omega} \mathcal{T}(\alpha, \beta, \gamma) d\bm{x}\,.
\end{equation}
The interaction effects between modes $1$ to $120$ is plotted in Fig.~\ref{fig:triad3}. The interaction of the first eight modes contributes to the net energy gain over the convection effect of all the modes, while the interaction of high order modes shows a net energy loss. The interaction of neighboring modes in the low order region $\alpha<20$ results in net energy gains, and the long-distance interaction between modes generally results in net energy losses from the convection effect of all modes. Non-local interactions exist primarily in the region of $\alpha\in [36, 46]$, where off-diagonal points represent non-local interactions. This means that low-order contributions are significant, mid-order contributions are weak, and the non-local contributions, which are mainly the interaction between low-order modes and high-order modes, are non-negligible. This result shows an agreement with stationary cavity flow results by Rubini et al.\cite{Rubini2020, Rubini2022}, which use a sparsification of Galerkin models to analyse energy interactions.

(2) {\it Convection effect of mode $\gamma$ on the interactions of mode $\alpha$ and $\beta$.} This contribution was studied by Couplet et al.\cite{Couplet2003} in a stationary pipe flow with a step. The triadic term $\mathcal{T}(\alpha, \beta, \gamma)$ is regarded as the convection contribution of mode $\gamma$ on the interactions of the mode pair, $(\alpha,\beta)$. The convection effect on the energy interaction is then
\begin{equation}\label{eq:pi}
    \Pi(\gamma\vert\alpha)=\sum_{\beta} \int_{\Omega} \mathcal{T}(\alpha, \beta, \gamma) d\bm{x}\,.
\end{equation}
The direction of energy transfer is recovered by the sign of the integral of $\Pi(\gamma\vert\alpha)$ in Fig.~\ref{fig:triad2}. The blue regions correspond to negative energy contribution that shares the same negative energy contribution transfer, i.e., to a net drain of the kinetic energy of the mode pair $(\gamma, \alpha)$. Moreover, the red region corresponds to positive energy, i.e., to a net gain of the kinetic energy of $(\gamma, \alpha)$. Since the first $106$ modes contribute to $90\%$ of the TKE and modes $107-438$ contribute only to $8\%$ of the TKE, we only plot Term III within the first set of modes to attain a clear interpretation of the high energy interactions. In the lower $\alpha$ mode region ($\alpha<10$), modes $\gamma<5$ and $\gamma>15$ are seen to lose energy, while modes $5<\gamma<15$ experience a gain. In the middle $\alpha$ mode region ($5<\alpha\le 15$), the majority of all modes shows energy loss. In the higher modes region $(\gamma\ge15, \alpha\ge15)$, the mode pair $(\gamma, \alpha)$ mostly loses energy.

Comparing to the study by Couplet et al.\cite{Couplet2003}, where $\gamma$ modes drain energy from modes $\alpha<\gamma$ and the non-local interactions are negligible, Fig.~\ref{fig:triad2} shows the middle region of $\gamma$ modes gain energy, and the low order $\gamma$ modes and high order $\gamma$ modes drain energy. Non-local interactions can be found in the range of $\alpha \in [20, 50]$.

\subsection{Non-local interaction and dynamics of energy transfers.}\label{sec:non-local}
\begin{figure*}[ht]
\begin{subfigure}[b]{0.46\textwidth}
        \includegraphics[width=\textwidth]{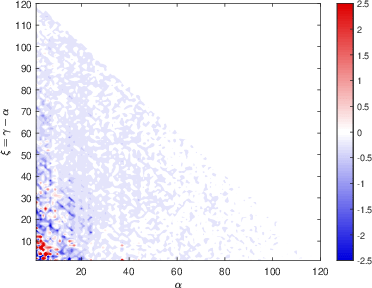}
        \caption{\label{fig:triad4}}
    \end{subfigure}
    \hfill
    \begin{subfigure}[b]{0.46\textwidth}
        \includegraphics[width=\textwidth]{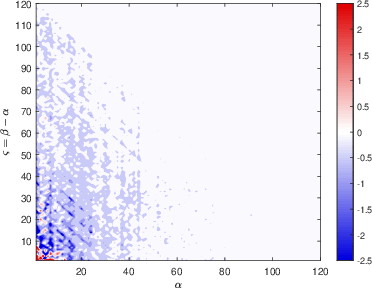}
        \caption{\label{fig:triad5}}
    \end{subfigure}
    \begin{subfigure}[b]{0.46\textwidth}
        \includegraphics[width=\textwidth]{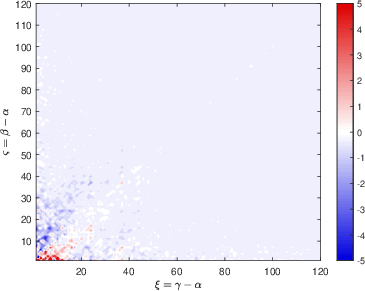}
        \caption{\label{fig:triad6}}
    \end{subfigure}

    \caption{Distribution of (\protect\subref{fig:triad4}) $\Theta_1(\alpha,\xi)$, (\protect\subref{fig:triad5}) $\Theta_2(\alpha,\varsigma)$ and (\protect\subref{fig:triad6}) $\Theta_3(\xi,\varsigma)$. There is clear non-negligible contribution in the non-local range of $\xi\ge25$ and $\varsigma\ge 25$. The color coding is concurrent with that of Fig.~\ref{fig:triad0}.}
\end{figure*}

Fig.~\ref{fig:pixi} shows that non-local interactions in the off-diagonal points are not negligible. It is thus necessary to investigate the interaction distance in Eq.~\eqref{eq:xi} and \eqref{eq:pi}, which is why two new functions with two distance variables are defined to find the relation between interaction distance and mode $\alpha$,
\begin{equation} \label{eq:theta}
\begin{cases}
    \Theta_1(\alpha,\xi)=\displaystyle\sum_{\beta =1} \int_{\Omega} \mathcal{T}(\alpha, \beta, \alpha+\xi) d\bm{x}\\[2ex]
    \Theta_2(\alpha,\varsigma)=\displaystyle\sum_{\gamma =1} \int_{\Omega} \mathcal{T}(\alpha, \alpha+\varsigma, \gamma) d\bm{x}
\end{cases},
\end{equation}
where the distance variable $\xi=\gamma -\alpha$ denotes the convection effect distance, while the distance variable $\varsigma=\beta-\alpha$ represents the interaction distance. $\Theta_1(\alpha,\xi)$ represents the distance formulation of function $\Pi(\gamma\vert \alpha)$ with the new distance variable $\xi=\gamma -\alpha$. The results are shown in Fig.~\ref{fig:triad4}.

It is seen that the primary energy transport occurs in the distance range of 20 modes. The local convection effect with a distance of less than 25 ($\xi< 25$) contribute to the main energy transport behavior. And the non-local convection effect with distance ($\xi\ge 25$) also shows clear evidence of energy transport. In the lower order modes region $\alpha \le 3$, the convection effect of short distance ($\xi<3$) shows energy loss and the convection effect of middle distance ($3\le\xi\le 13$) shows a net gain of energy. The convection effect of long distances ($\xi\ge 14$) depicts energy losses for most of the mode pairs. $\Theta_2(\alpha,\varsigma)$ represents the distance formulation of function $\Xi(\alpha,\beta)$ with a new distance variable $\varsigma=\gamma -\alpha$. In Fig.~\ref{fig:triad5}, the dominating energy transport occurs within the distance range of 25 modes. In the lower-order modes region $\alpha \le 15$, the main interaction effect shows net energy gain in the near distance range $\varsigma\le 3$, and the non-local interaction effect with distance ($\xi\ge 25$) is not negligible.

To distinguish the convection effect from the interaction effect due to non-local transfer, a new function $\Theta_3(\xi,\varsigma)$ can be applied with distance variables $\xi$ and $\varsigma$ in Eq.~\eqref{eq:theta},
\begin{equation}
    \Theta_3(\xi,\varsigma)=\sum_{\alpha =1} \int_{\Omega} \mathcal{T}(\alpha, \alpha+\varsigma, \alpha+\xi) d\bm{x}\,.
\end{equation}

The results are plotted in Fig.~\ref{fig:triad6}, which shows a disagreement with Couplet et al.\cite{Couplet2003}. The energy transfer among the Phase POD modes includes non-local transfer, as seen from the non-negligible energy transfer in the region $\xi\ge 25$ and $\varsigma\ge 25$. However, in the local transfer region $\xi < 25$ and $\varsigma < 25$, the convection effect where $\xi\ge\varsigma$ shows a net gain of energy, and the interaction convection effect where $\xi<\varsigma$ shows net loss of energy. The stationary example in Couplet et al.\cite{Couplet2003} shows a local energy transfer process, so we can assume that the non-stationarity contributes to the non-local energy transfer.

\begin{figure*}
\begin{subfigure}[b]{0.46\textwidth}
        \includegraphics[width=\textwidth]{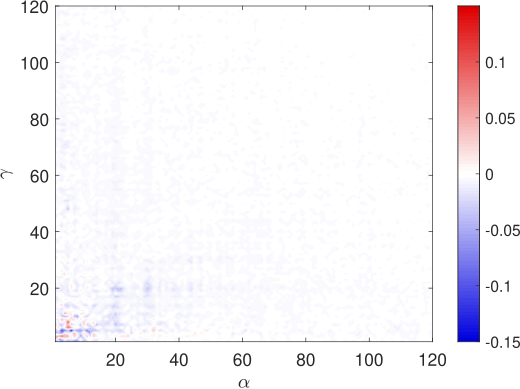}
        \caption{\label{fig:Pi10}}
    \end{subfigure}
    \hfill
    \begin{subfigure}[b]{0.46\textwidth}
        \includegraphics[width=\textwidth]{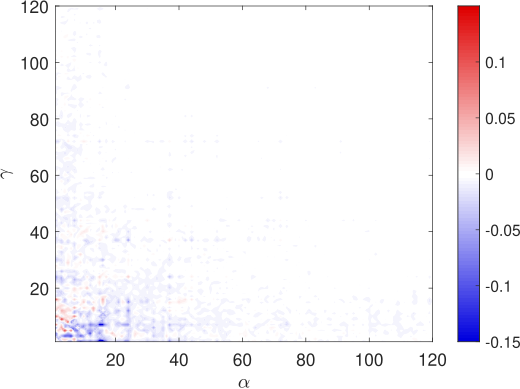}
        \caption{\label{fig:Pi60}}
    \end{subfigure}

    \begin{subfigure}[b]{0.46\textwidth}
        \includegraphics[width=\textwidth]{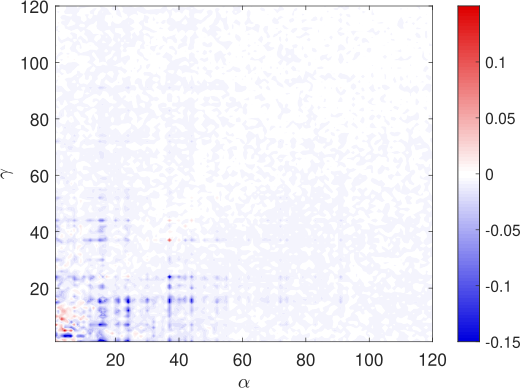}
        \caption{\label{fig:Pi80}}
    \end{subfigure}
    \hfill
    \begin{subfigure}[b]{0.46\textwidth}
        \includegraphics[width=\textwidth]{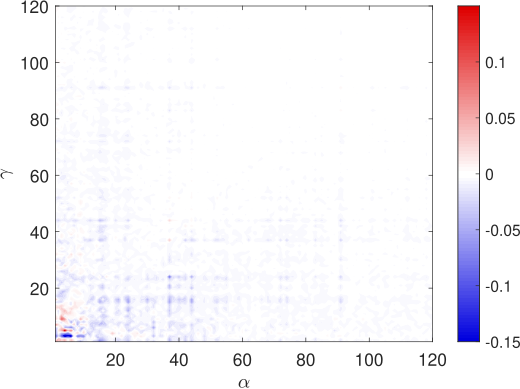}
        \caption{\label{fig:Pi95}}
    \end{subfigure}
    
    \caption{Dynamics of the triadic interaction $\Pi_p$ for $p=\{0.1T, 0.6T,0.8T,0.95T\}$. The color coding is concurrent with that of Fig.~\ref{fig:triad0}. In (\protect \subref{fig:Pi10}) $\Pi_p(\gamma\vert \alpha,p=0.1T)$, local energy dominates the triadic interactions. (\protect \subref{fig:Pi60}) $\Pi_p(\gamma\vert \alpha,p=0.6T)$, non-local energy transfer begins to emerge. (\protect \subref{fig:Pi80}) $\Pi_p(\gamma\vert \alpha,p=0.8T)$, pronounced non-local energy transfer. (\protect \subref{fig:Pi95}) $\Pi_p(\gamma\vert \alpha,p=0.95T)$, return to predominating local energy transfer. The videos view by: https://doi.org/10.11583/DTU.21444684.\cite{Zhang2022}}
    \label{fig:triadicdynamics0}
\end{figure*}

For the study of non-local interactions in the non-stationary case, we consider only the spatial domain $\Omega_S$ in the integral of interaction terms in Eq.~\eqref{eq:xi} and \eqref{eq:pi}
\begin{equation}\label{eqn:intterms}
\begin{cases}
\Pi_p(\gamma\vert\alpha,p)=\displaystyle\sum_{\beta} \int_{\Omega_S} \mathcal{T}(\alpha, \beta, \gamma) d\bm{x}\\[2ex]
\Xi_p(\alpha,\beta,p)=\displaystyle\sum_{\gamma} \int_{\Omega_S} \mathcal{T}(\alpha, \beta, \gamma) d\bm{x}
\end{cases}.
\end{equation}

\begin{figure*}
    \begin{subfigure}[b]{0.46\textwidth}
        \includegraphics[width=\textwidth]{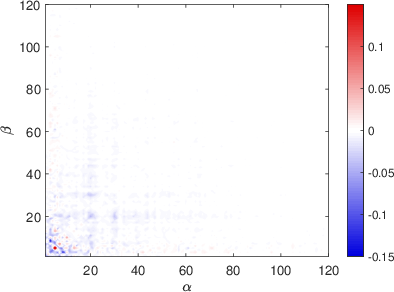}
        \caption{\label{fig:Xi10}}
    \end{subfigure}
    \hfill
    \begin{subfigure}[b]{0.46\textwidth}
        \includegraphics[width=\textwidth]{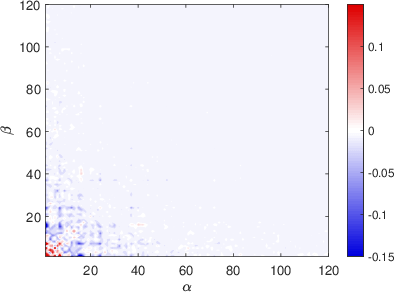}
        \caption{\label{fig:Xi60}}
    \end{subfigure}

    \begin{subfigure}[b]{0.46\textwidth}
        \includegraphics[width=\textwidth]{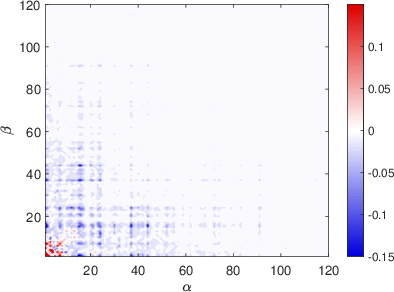}
        \caption{\label{fig:Xi80}}
    \end{subfigure}
    \hfill
    \begin{subfigure}[b]{0.46\textwidth}
        \includegraphics[width=\textwidth]{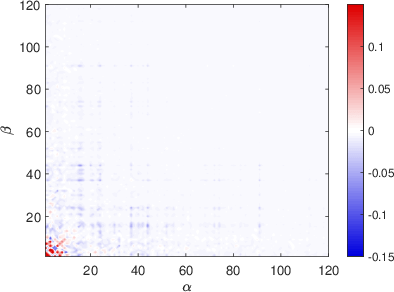}
        \caption{\label{fig:Xi95}}
    \end{subfigure}

    \caption{Dynamics of the triadic interaction $\Xi_p$ for $p=\{0.1T, 0.6T,0.8T,0.95T\}$. The color coding is concurrent with that of Fig.~\ref{fig:triad0}. In (\protect \subref{fig:Xi10}) $\Xi_p(\alpha,\beta, p=0.1T)$, local energy dominates the triadic interactions. (\protect \subref{fig:Xi60}) $\Xi_p(\alpha,\beta,p=0.6T)$, non-local energy transfer begins to emerge. (\protect \subref{fig:Xi80}) $\Xi_p(\alpha,\beta,p=0.8T)$, pronounced non-local energy transfer. (\protect \subref{fig:Xi95}) $\Xi_p(\alpha,\beta,p=0.95T)$, return to predominating local energy transfer. The videos view by: https://doi.org/10.11583/DTU.21444684.\cite{Zhang2022}}
    \label{fig:triadicdynamics1}
\end{figure*}

Then the triadic interaction term can be evaluated as a function of phase to evaluate the non-stationary effects on the mode relations and interactions. A movie of $\Pi_p(\gamma\vert\alpha,p)$ and $\Theta_p(\alpha, \beta, p)$ is available at \url{https://doi.org/10.11583/DTU.21444684} \cite{Zhang2022} and on the web page of the DTU Turbulence Research Lab. Four snapshots of both interaction terms (Eq.~\ref{eqn:intterms}) for $p=\{0.1T, 0.6T,0.8T,0.95T\}$ are shown in Fig.~\ref{fig:triadicdynamics0} and \ref{fig:triadicdynamics1}. At the beginning of the period (Fig.~\ref{fig:Pi10} and Fig.~\ref{fig:Xi10}), energy transfer occurs mainly between the low-order modes (mode order less than 20). The energy transfer is mainly local. From $p=0.6T$ (Fig.~\ref{fig:Pi60} and Fig.~\ref{fig:Xi60}), non-local energy transfer begins to increasingly emerge. At $p=0.8T$ in Fig.~\ref{fig:Pi80} and Fig.~\ref{fig:Xi80}, pronounced non-local energy transfer emerges in the middle-order modes (mode order from $25$ to $50$). Although the energy is very low compared to the non-local energy transfer of low-order modes, it is significantly more apparent in comparison to the energy transfer prior to $p=0.8T$. At the end of the period in Fig.~\ref{fig:Pi95} and Fig.~\ref{fig:Xi95}, the energy transfer returns to primarily local interactions as in the beginning of the period. From the best knowledge of the authors, the mechanisms behind the triadic interaction corresponding to the phase varying are still unknown, which is also not a central focus point in this paper, but the Phase POD can contribute to the data analysis necessary to study these kinds of non-stationary mechanisms.

For a view of the relative contributions of local and non-local transfer, we separate the modal elements of $\Pi_p(\gamma\vert \alpha)$ and $\Xi_p(\alpha, \beta)$ into two groups -- the local group and the non-local group. Couplet et al.\cite{Couplet2003} chose $25$ as the distance limit characterizing locality, in order to obtain local transfer conclusion with the POD modes. We also use 25 as the local distance range. The local convection effect $\Pi_l(p)$ and the non-local convection effect $\Pi_n(p)$ are chosen as
\begin{equation}
\begin{cases}
\Pi_l(p)=\displaystyle\sum_{\alpha,\gamma} \big\vert \Pi_p(\gamma\vert \alpha,p) \big\vert, \mbox{ where } \vert\alpha -\gamma\vert <25 \\[2ex]
\Pi_n(p)=\displaystyle\sum_{\alpha,\gamma} \big\vert \Pi_p(\gamma\vert \alpha,p) \big\vert, \mbox{ where } \vert\alpha -\gamma\vert \ge 25\\[2ex]
\end{cases}.
\label{eq:pin}
\end{equation}
So are the local interaction effect $\Xi_l(p)$ and the non-local interaction effect $\Xi_n(p)$:
\begin{equation}
\begin{cases}
\Xi_l(p)=\displaystyle\sum_{\alpha,\beta} \big\vert \Xi_p(\alpha, \beta, p) \big\vert, \mbox{ where } \vert\alpha -\beta\vert <25 \\[2ex]
\Xi_n(p)=\displaystyle\sum_{\alpha,\beta} \big\vert \Xi_p(\alpha, \beta, p) \big\vert, \mbox{ where } \vert\alpha -\beta\vert \ge 25\\[2ex]
\end{cases}.
\label{eq:xin}
\end{equation}

\begin{figure*}[ht]
        \hspace{0.2cm}
        \begin{subfigure}[b]{0.44\textwidth}
        \includegraphics[width=\textwidth]{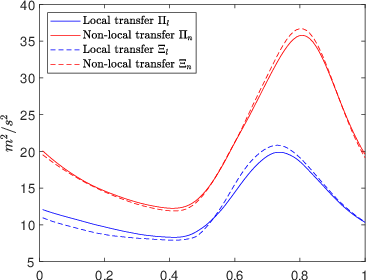}
        \caption{\label{fig:nonlocal25}}
        \end{subfigure}
       \hfill
        \begin{subfigure}[b]{0.46\textwidth}
        \includegraphics[width=\textwidth]{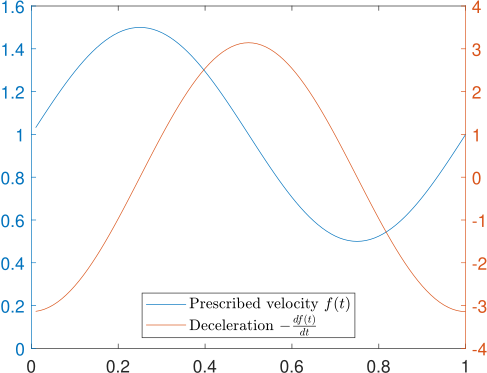}
        \caption{\label{fig:ft}}
        \end{subfigure}
        
        \begin{subfigure}[b]{0.44\textwidth}
        \includegraphics[width=\textwidth]{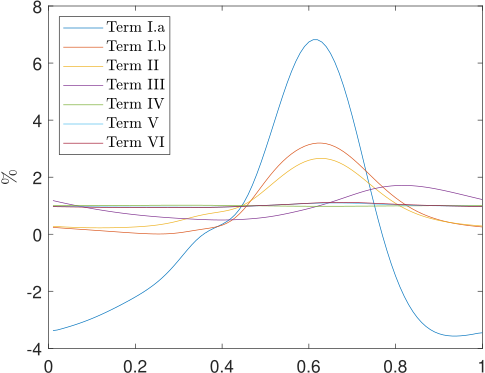}
        \caption{\label{fig:contributions}}
        \end{subfigure}
       \hfill
        \begin{subfigure}[b]{0.46\textwidth}
        \includegraphics[width=\textwidth]{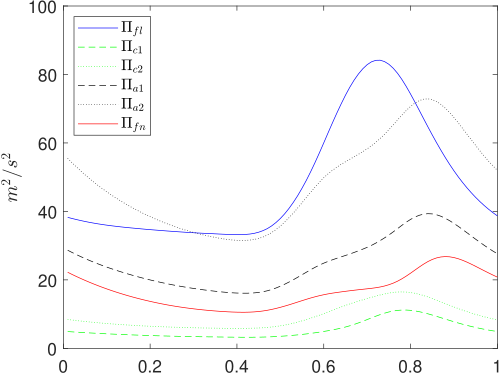}
        \caption{\label{fig:nonlocal2k}}
        \end{subfigure}

    \caption{Energy transfer in one period. The x-axis in (\protect\subref{fig:nonlocal25}), (\protect\subref{fig:ft}) and (\protect\subref{fig:nonlocal2k}) denotes phases in $(0, T]$. (\protect\subref{fig:nonlocal25}) The local and non-local energy transfer contribution of $\Pi_l(p)$, $\Pi_n(p)$, $\Xi_l(p)$ and $\Xi_n(p)$. (\protect\subref{fig:ft}) The prescribed velocity $f(t)$ and deceleration $-df(t)/dt$. (\protect\subref{fig:contributions}) Contributions of all terms in equation \ref{eq:TKEequation}, each term is integral over space and normalized by themselves with integral over space and time. (\protect\subref{fig:nonlocal2k}) The local, half-local and non-local energy transfer of $\Pi_{fl}$, $\Pi_{c1}$, $\Pi_{c2}$, $\Pi_{a1}$, $\Pi_{a2}$ and $\Pi_{fn}$ as $BW=25$.}
\end{figure*}
The results of implementing Eq.~\eqref{eq:pin} and \eqref{eq:xin} on the non-stationary turbulent data are shown in Fig.~\ref{fig:nonlocal25}. It can be seen that, in general, both the non-local convection effect $\Pi_n(p)$ and the non-local interaction effect $\Xi_n(p)$ contribute more than the local energy transfer. In this statistically non-stationary turbulent flow, the non-local energy transfer is the predominating type of mechanism for energy transfer. Both the local and, in particular, non-local energy transfer increase dramatically from about $0.5\,T$, when the periodic deceleration of the lid, $-df(t)/dt$, reaches its peak (Fig.~\ref{fig:ft}). The triadic interactions are thus highly correlated with the deceleration of the lid, which naturally also directly affects the Term I.a in Eq.~\eqref{eq:TKEequation} at the unsteady top wall. 

For a more detailed breakdown of the transport of energy from the non-stationary top wall (boundary conditions Eq.~\eqref{eq:lid_velocity} and~\eqref{eq:ft}) to the cavity flow, we computed all terms in Eq.~\eqref{eq:TKEequation} by integrating each term across space and normalizing by the corresponding integral over space and time. As shown in Fig.~\ref{fig:contributions}, it is seen that the rate of change of TKE (Term I.a) responds firstly to the deceleration function with a delay of approximately $0.05\,T$. The local energy transfer of Term III responds to Term I.a with a $0.2\,T$ delay, and the non-local energy transfer shows a delay of around $0.05\,T$ from the local energy transfer in both $\Pi_n$ and $\Xi_n$. Kraichnan\cite{Kraichnan1971, Kraichnan1976} presented a result based on Fourier modes that wavenumber $k$ will exchange most of its energy with modes within the range $[k/2, 2k]$, i.e. that is within the local transfer range setting. We computed $\Pi_l(p)$, $\Pi_n(p)$, $\Xi_l(p)$ and $\Xi_n(p)$ with the local range $[\alpha/2, 2\alpha]$, and the non-local energy transfer shows higher values than the one with local range $[\alpha-25, \alpha+25]$ due to the narrow range in the low-order $\alpha$ modes.

\begin{table*}[ht]

\begin{ruledtabular}
  \begin{tabular}{clcccc}
     \textbf{Domain no.} & \textbf{Defined domain } & $\vert \alpha -\beta \vert$   &   $\vert \alpha - \gamma \vert$ & $\vert \beta - \gamma \vert$ & \textbf{symbols } \\[4pt]
      \hline
       \multirow{1}{2cm}{(1)}&\multirow{1}{8cm}{Fully local convection on local interaction (FCFI)} & $<BW$ & $<BW$ & $<BW$ & $\Pi_{fl}$\\[4pt]
       \hline
       \multirow{1}{2cm}{(2.a)}&\multirow{1}{8cm}{Local convection on non-local interaction (FCNI)}& $\ge BW$ & $< BW$ & $<BW$ & $\Pi_{c1}$\\[4pt]
       \hline
       \multirow{2}{2cm}{(2.b)}&\multirow{2}{8cm}{Half non-local convection on local interaction (HCFI)}&$<BW$&$\ge BW$&$<BW$& \multirow{2}{*}{$\Pi_{c2}$ }\\
        & &$<BW$&$< BW$&$\ge BW$& \\[4pt]
         \hline
        \multirow{1}{2cm}{(3.a)}&\multirow{1}{8cm}{Fully non-local convection on local interaction (NCFI)}& $<BW$ & $\ge BW$ & $\ge BW$ & $\Pi_{a1}$\\[4pt]
        \hline
        \multirow{2}{2cm}{(3.b)}& \multirow{2}{8cm}{Half non-local convection on non-local interaction (HCNI)}   & $\ge BW$ & $< BW$ & $\ge BW$ & \multirow{2}{*}{$\Pi_{a2}$ }\\
        & &$\ge BW$&$\ge BW$&$< BW$& \\[4pt]
         \hline
       \multirow{1}{2cm}{(4)}&\multirow{1}{8cm}{Non-local convection on non-local interaction (NCNI)}& $\ge BW$ & $\ge BW$ & $\ge BW$ & $\Pi_{fn}$\\[4pt]
  \end{tabular}
  \end{ruledtabular}
  \caption{\label{tab:range} Local, half non-local and non-local ranges with the bandwith $BW$ regarding to Eq.~\eqref{eq:gamma}}
\end{table*}

Because $\Pi_p(\gamma\vert\alpha)$ represents the sum of energy transfer of all the modes $\beta$, including the local and non-local energy interactions, we cannot use two regions to define the local and non-local energy transfer with two distance variables between three modes number. Firstly, three distance variables should be used to describe the distances among the three mode variables, $(\alpha, \beta, \gamma)$. Secondly, we defined the bandwidth of locality $BW$ to distinguish the local energy transfer range and the non-local energy transfer range among these three modes. Then we define six new functions with three distance variables to express fully local, half-local and fully non-local energy transfer:
\begin{widetext}
\begin{equation}
\begin{cases}
\Pi_{fl}(p)=\displaystyle\sum_{\alpha,\beta,\gamma} \big\vert \int_{\Omega_S} \mathcal{T}(\alpha, \beta, \gamma) d\bm{x} \big\vert, \mbox{ where $(\alpha, \beta, \gamma)$ is in domain (1) in table \ref{tab:range}} \\[2ex]
\Pi_{c1}(p)=\displaystyle\sum_{\alpha,\beta,\gamma} \big\vert\int_{\Omega_S} \mathcal{T}(\alpha, \beta, \gamma) d\bm{x} \big\vert, \mbox{ where $(\alpha, \beta, \gamma)$ is in domain (2.a) in table \ref{tab:range}} \\[2ex]
\Pi_{c2}(p)=\displaystyle\sum_{\alpha,\beta,\gamma} \big\vert\int_{\Omega_S} \mathcal{T}(\alpha, \beta, \gamma) d\bm{x} \big\vert, \mbox{ where $(\alpha, \beta, \gamma)$ is in domain (2.b) in table \ref{tab:range}} \\[2ex]
\Pi_{a1}(p)=\displaystyle\sum_{\alpha,\beta,\gamma} \big\vert\int_{\Omega_S} \mathcal{T}(\alpha, \beta, \gamma) d\bm{x} \big\vert, \mbox{ where $(\alpha, \beta, \gamma)$ is in domain (3.a) in table \ref{tab:range}} \\[2ex]
\Pi_{a2}(p)=\displaystyle\sum_{\alpha,\beta,\gamma} \big\vert\int_{\Omega_S} \mathcal{T}(\alpha, \beta, \gamma) d\bm{x} \big\vert, \mbox{ where $(\alpha, \beta, \gamma)$ is in domain (3.b) in table \ref{tab:range}} \\[2ex]
\Pi_{fn}(p)=\displaystyle\sum_{\alpha,\beta,\gamma} \big\vert\int_{\Omega_S} \mathcal{T}(\alpha, \beta, \gamma) d\bm{x} \big\vert, \mbox{ where $(\alpha, \beta, \gamma)$ is in domain (4) in table \ref{tab:range}} 
\end{cases},
\label{eq:gamma}
\end{equation}
\end{widetext}
where $\Pi_{fl}(p)$ denotes the fully local energy transfer within the local distance is set as $BW$ in Table~\ref{tab:range}, $\Pi_{c1}(p)$ denotes the fully local convection effect on the non-local interaction, $\Pi_{c2}(p)$ denotes the half local convection effect on the local interaction, $\Pi_{a1}(p)$ denotes the fully non-local convection effect on the local interaction, $\Pi_{a2}(p)$ denotes the half non-local convection effect on the non-local interaction, and $\Pi_{fn}(p)$ denotes the fully non-local energy transfer. In the same manner as the local energy transfer observed in Couplet et al.\cite{Couplet2003}, we have likewise set $BW=25$ and shown the results for Eq.~\eqref{eq:gamma} in Fig.~\ref{fig:nonlocal2k}. It can be seen that the fully local energy transfer $\Pi_{fl}(p)$ contributes significantly among all six functions. However, the half non-local convection term $\Pi_{a2}(p)$ shows a similar behavior as the fully local energy transfer $\Pi_{fl}(p)$. The fully non-local convection term $\Pi_{a1}(p)$ also shows strong and similar impact as the full non-local energy transfer term $\Pi_{fn}(p)$. 

In contrast, the local convection on non-local interaction $\Pi_{c1}$ and the half non-local convection on local interaction $\Pi_{c2}$ show weak energy exchanges even when there are two distance variables in the local domain. The local energy exchange mainly occurs in the fully local domain. The non-local energy exchange is composed of the non-local convection on non-local interactions $\Pi_{a2}$, the fully non-local convection on the local interaction term $\Pi_{a1}$ and the full non-local energy transfer term $\Pi_{fn}$, of which there are at least two non-local distances among the three distance variables.

\begin{figure*}[ht]
        \begin{subfigure}[b]{0.46\textwidth}
        \includegraphics[width=\textwidth]{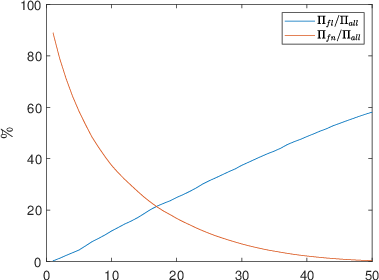}
        \caption{\label{fig:nonlocalphase}}
        \end{subfigure}
        \begin{subfigure}[b]{0.46\textwidth}
        \includegraphics[width=\textwidth]{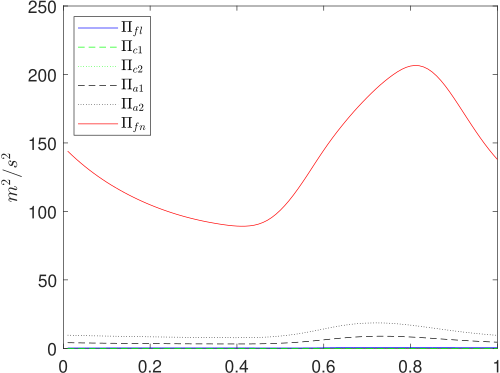}
        \caption{\label{fig:nonlocal1}}
        \end{subfigure}
        
        \hspace{0.2cm}
        \begin{subfigure}[b]{0.46\textwidth}
        \includegraphics[width=\textwidth]{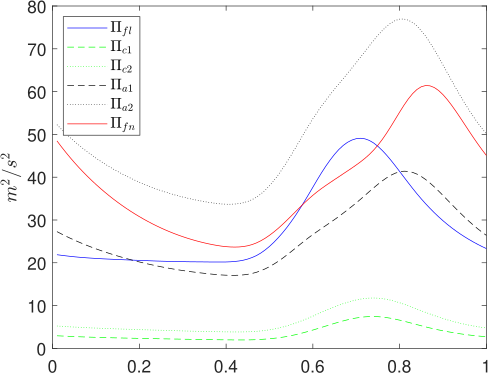}
        \caption{\label{fig:nonlocal15}}
        \end{subfigure}
        \begin{subfigure}[b]{0.46\textwidth}
        \includegraphics[width=\textwidth]{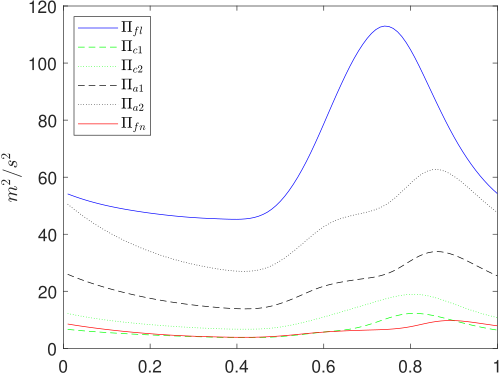}
        \caption{\label{fig:nonlocal35}}
        \end{subfigure}

    \caption{The local and non-local energy transfer contributions with different bandwidths. The x-axis in (\protect\subref{fig:nonlocal1}), (\protect\subref{fig:nonlocal15}) and (\protect\subref{fig:nonlocal35}) denotes phases in $(0, T]$. (\protect\subref{fig:nonlocalphase}) Contributions of fully local transfer and fully non-local transfer as the bandwidth $BW$ in x-axis increase from 1 to 50, all the terms in \eqref{eq:gamma} are normalized by $\Pi_{all}=\Pi_{l}+\Pi_{c1}+\Pi_{c2}+\Pi_{a1}+\Pi_{a2}+\Pi_{n}$. (\protect\subref{fig:nonlocal1}) The local and non-local energy transfer contributions for $BW=1$. (\protect\subref{fig:nonlocal15}) The local and non-local energy transfer contributions for $BW=15$. (\protect\subref{fig:nonlocal35}) The local and non-local energy transfer contributions for $BW=35$.}
\end{figure*}

The bandwidth of locality is the key factor in discerning between local and non-local energy transfer. As shown in Fig.~\ref{fig:nonlocalphase}, increasing the bandwidth of locality can include more local energy transfer from the non-local domain. For the example of $BW=1$ in Fig.~\ref{fig:nonlocal1}, where the local energy transfer only present TKE self-convection, $\mathcal{T}(\alpha, \alpha,\alpha) = -\frac{1}{2}\langle a_\alpha a^{\alpha*} a_\alpha\rangle\varphi^j_\alpha \nabla_j \bm{k}^\alpha_\alpha$, the fully local transfer contribution $\Pi_{fl}$ is negligible compared with the fully non-local energy transfer $\Pi_{fn}$. In this case, the terms $\Pi_{c1}$ and $\Pi_{c2}$ do not exist because there is no solution for the inequalities of the domains in Table~\ref{tab:range}. As the bandwidth increases, the fully non-local energy transfer $\Pi_{fn}$ is converted to the fully non-local convection on local interaction term $\Pi_{a1}$ and the half non-local convection on non-local interaction $\Pi_{a2}$. The non-local terms $\Pi_n$ and $\Xi_n$ in Fig.~\ref{fig:nonlocal25} consist of the full non-local energy transfer term, the fully non-local convection on local interaction term $\Pi_{a1}$ and the half non-local convection on non-local interaction  $\Pi_{a2}$. For this reason, both of the non-local terms $\Pi_n$ and $\Xi_n$ contribute more than the local energy transfer in Fig.~\ref{fig:nonlocal25}.

Furthermore, a significant time delay pattern from local to non-local transfer can be found in Fig.~\ref{fig:nonlocal25}, \ref{fig:nonlocal2k}, \ref{fig:nonlocal15} and \ref{fig:nonlocal35}. The peak of the fully local energy transfer occurs initially, and is followed by the local convection on non-local interactions $\Pi_{c1}$ and the half non-local convection on local interaction $\Pi_{c2}$ with $0.06\,T$ delay, which have two distance variable in the local domain. The peaks of the fully non-local convection on local interaction term $\Pi_{a1}$ and the half non-local convection on non-local interaction $\Pi_{a2}$ also show a $0.06\,T$ delay comparing with $\Pi_{c1}$ and $\Pi_{c2}$ The fully non-local term $\Pi_{fn}$ reaches its peak $0.06T$ later than the ones in $\Pi_{a1}$ and $\Pi_{a2}$.

In summary, the non-local energy transfer shows non-negligible contributions in the energy transfer process. The non-stationarity (especially the non-stationary term Term I.a) delivers energy to all the local, half-local, and non-local terms but with varying time delays. The bandwidth of the locality is the key factor in deciding the local energy proportion due to very weak TKE self-convection effect. The temporal delays between the non-local terms in Table~\ref{tab:range} are observed, but the mechanism behind this is still unknown. The Phase POD provides a new approach to studying the non-stationary triadic interactions of energy optimized modes.


\section{Conclusions}\label{sec:conclusion}
We demonstrate a direct temporal-extension POD method (Phase POD) with phase averaging for describing the four-dimensional (spatial and temporal) behaviour in statistically non-stationary turbulent flows. As Phase POD includes all temporal-spatial information in the modes, the full-dimensional coherent structure can be extracted from non-stationary turbulent flows. Phase POD does not depend on the assumption of homogeneity in the time dimension. However, with some assumptions in the temporal dimension, space only POD and Spectral POD are special cases of Phase POD. The modes of Phase POD will have energy optimality, including temporal properties. This extension in the time domain subverted the concept of time-dependent coefficients, which is always used in Galerkin projection to analyze stationary energy transfer and reduced-order models. Instead, the coefficients in the Phase POD are the energetic weights of modes. 

Time averaging is not suitable for periodic or non-stationary flows. Phase averaging is a convenient manner to conduct the Phase POD. We show the equivalence of phase averaging and ensemble averaging, which follows a similar definition in the proof that time averaging equals ensemble averaging in a statistically steady flow. Also, the inequivalence of time average and phase average is illustrated with the probability conservation of the period-shift operator in a periodic flow case.

Although the Phase POD can resolve four-dimensional coherent structures, the size of the correlation matrix introduces a computational challenge. The snapshots method is formulated to convert an impractical sized matrix to a computationally practical size. Despite increasing the data reading time from $O(N)$ to $O(N^2)$, the snapshots method can handle the Phase POD with higher resolution in any coordinates.

We apply the Phase POD method to an example of a statistically non-stationary turbulent cavity flow and extract the spatio-temporal modes that describe the complete turbulent flow dynamics. Four patterns are found to show how the energy structures evolve with phase. By substituting the velocity variables with the modes in the TKE equation, the contribution of each mode is shown as a function of phase. The triadic interaction term (Term III) can be interpreted as the convection effect from the interaction of two modes. Local energy transfer dominates the main energy transfer behaviour. However, non-local energy transfer is detected in this statistically non-stationary cavity flow. In the dynamics analysis of the triadic interaction, non-stationarity is confirmed to trigger the non-local interactions.


\begin{acknowledgments}
Yisheng Zhang and Azur Hod\v zi\'c acknowledge financial support from the European Research council: This project has received funding from the European Research Council (ERC) under the European Unions Horizon 2020 research and innovation program (grant agreement No 803419).

Fabien Evrard has received funding from the European Union’s Horizon 2020 research and innovation program under the Marie Skłodowska-Curie Grant Agreement No. 101026017. 

Clara M. Velte acknowledges financial support from the Poul Due Jensen Foundation: Financial support from the Poul Due Jensen Foundation (Grundfos Foundation) for this research is gratefully acknowledged.
\end{acknowledgments}


\section*{Data Availability Statement}
Raw data are available from the corresponding author upon reasonable request. The data and the Matlab code that support the findings of this study are openly available at \url{https://doi.org/10.11583/DTU.21444684}\cite{Zhang2022}.
\section*{Author Declarations}
The authors have no conflicts to disclose.


\appendix
\section{Equivalence relation between phase and ensemble averaging in periodically driven flows}\label{appA}

Three definitions for averaging are mainly used with empirical data: ensemble averaging, time averaging and phase averaging. Time averaging is equivalent to ensemble averaging in stationary flows with the assumption of probability conservation under any time-translation in stationary flows (see p. 100 in Ref.~\onlinecite{Holmes2012}, and p. 47 in Ref.~\onlinecite{Frisch1995}). Phase averaging is popular in applications of periodic flows. This section discusses the relation between phase averaging and ensemble averaging in  periodically driven flows.

\subsection{Dynamic system of incompressible periodically driven flow} 
Incompressible periodically driven flow is governed by the non-dimensional Navier-Stokes equation
\begin{equation}\label{eq:NS}
\left\{ \begin{array}{rll}
\displaystyle \mathrm{Str}\frac{\partial \tilde{\bm{u}}}{\partial t} + \tilde{\bm{u}}\cdot \nabla\tilde{\bm{u}}  =& \displaystyle-\nabla p +\frac{1}{\mathrm{Re}}\nabla^2\tilde{\bm{u}}\\[8pt]
\nabla\cdot \tilde{\bm{u}} =&0\\[8pt]
        \tilde{\bm{u}}_0 \equiv&\tilde{\bm{u}}(\bm{r},p_0)& \mbox{initial condition } (t=0) \\[8pt]
        \tilde{\bm{u}}(\bm{r}_0,t+p_0) =&\tilde{\bm{u}}(\bm{r}_0,t+p_0+T)& \mbox{boundary condition}
 \end{array}\right.,
\end{equation}
where $\mathrm{Str}$ is the Strouhal number, $\mathrm{Re}$ is the Reynolds number, $p_0$ denotes the start phase in temporal dimension, $\bm{r}_0$ is the spatial boundary condition, and $\bm{r}$ is the spatial variable. Eq.~\eqref{eq:NS} can be considered a dynamic system that processes initial conditions and periodic boundary conditions of a velocity field at any time $t$.

A dynamic system $(\mathcal{H},\mathcal{A},P,G_t)$ is defined to describe a stochastic process based on the incompressible Navier-Stokes equation \cite{Holmes2012, Frisch1995}. The set $\mathcal{A}$ is a measurable space in $\mathcal{H}$. $P$ is the probability measure on $\mathcal{H}$. The time-shift operator $G_t$ is the mapping operator from both initial conditions and boundary conditions to the velocity vector at time $t$ by the Navier-Stokes equation \eqref{eq:NS}. So, for any time $t$, the velocity $\tilde{\bm{u}}(\bm{r},t)$ is generated as the operator $G_t$ maps the initial condition and boundary condition to a velocity at time $t$, $\tilde{\bm{u}}(\bm{r},t)=G_t(\tilde{\bm{u}}_0(\bm{r},p_0),\tilde{\bm{u}}(\bm{r}_0,p_0))$.

In stationary cases, an ansatz is stated that there exists a $G_t$ invariant probability measure $P$ on $\mathcal{H}$, for any subset $A$ in phase space satisfying $P(G_t^{-1}A)=P(A)$ \cite{Holmes2012, Frisch1995}, where $G_t^{-1}A$ denotes the set of initial conditions and boundary conditions which are mapped to $A$. In periodically driven fully developed turbulent flows, the ansatz can be restated as the periodic operator $G_T$ invariant probability measure $P$ on $\mathcal{H}$, for any subset $A$ in phase space satisfying the probability conservation:
\begin{equation}\label{eq:probabilityconservation}
     P(G_T^{-1}A)=P(A)
\end{equation}
where $G_T^{-1}A$ denotes the set of initial conditions and boundary conditions which are mapped onto $A$ after one period. Unlike the strict condition of time averaging that requires the probability conservation for any time-shifts in a stationary turbulent flow, a periodic turbulent flow only requires that each point with three spatial dimensions and one phase dimension shares the same probability distribution across periods. The period-shift operator $G_T$ also features the compact property of fully developed turbulence, in which the initial conditions, boundary conditions, and the velocity profile at any time developed from the initial conditions and boundary conditions are all in the same probability space.

\subsection{Ensemble averaging and properties of statistical moments of a periodically driven fully developed turbulent flow}
The ensemble average of a random velocity variable $\tilde{\bm{u}}$ is defined in the measurement space \citep{Frisch1995},
\begin{equation}\label{eq:ensembleaverage}
    \langle \tilde{\bm{u}} \rangle_e = \int_{\mathcal{H}} \tilde{\bm{u}} dP.
\end{equation}
So the ensemble measures the mean value of velocity vectors generated from all the possible initial conditions and boundary conditions. In a periodically turbulent flow, the ensemble average is based on a subset $Y_n\subset \mathcal{H}$ that includes all phases in a certain period (i.e. the $n$th period) as $Y_n: \Omega\times(nT,(n+1)T]\rightarrow V$. In this subset $Y_n$, the ensemble average can be written as
\[
\langle \tilde{\bm{u}} \rangle_e = \int_{Y_n} \tilde{\bm{u}}(\bm{r},p)dP = \lim_{N_e \to \infty} \frac{1}{N_e}\sum^{N_e}_{n_e=1}\tilde{\bm{u}}(\bm{r},p;n_e)
\]
where $(\bm{r},p)\in\Omega \times(nT,(n+1)T]$, $\tilde{\bm{u}}(\bm{r},p;n_e)\in Y_n$ denotes the $n_e$th repeat of the dynamics system, and $N_e$ is the total number of repetitions of the dynamic system.

Similar to the property of $G_t$ invariance in the stationary flow (see p. 47 in Ref.~\onlinecite{Frisch1995}), the moments of $G_T$ mapping a velocity variable in the periodical flow, if they exist, are invariant under the periodic-translation of all their temporal arguments:
\begin{eqnarray}\label{eq:ensemblemoments}
  &  \langle \tilde{\bm{u}}(\bm{r},p_1+T)\tilde{\bm{u}}(\bm{r}, p_2+T)\cdots \tilde{u}(\bm{r},p_m+T)\rangle_e \nonumber\\
  &=\langle \tilde{\bm{u}}(\bm{r},p_1)\tilde{\bm{u}}(\bm{r},p_2)\cdots \tilde{\bm{u}}(\bm{r},p_m)\rangle_e,
\end{eqnarray}
where $p_i\in (nT,(n+1)T]$ with $i\in \{1,2,\cdots,m\}$, a relation $p_1<p_2<\cdots<p_m$, and $\tilde{\bm{u}}(\bm{r},p_i)\in Y_n$. Eq.~\eqref{eq:ensemblemoments} declares that the moments under ensemble averaging are invariant over the period in a periodically driven fully developed turbulent flow. It is also easy to deduce that the moments of any integer $k$ ($\forall k\in \mathbb{N}$) period-translation are invariant:
\begin{eqnarray}
\label{eq:corollary}
&\langle \tilde{\bm{u}}(\bm{r},p_1')\tilde{\bm{u}}( \bm{r},p_2')\cdots \tilde{\bm{u}}(\bm{r},p_m')\rangle_e \nonumber\\
=&\langle \tilde{\bm{u}}(\bm{r},p_1)\tilde{\bm{u}}(\bm{r},p_2)\cdots \tilde{\bm{u}}(\bm{r},p_m)\rangle_e,
\end{eqnarray}
where $p_i'=p_i+kT$ is synchronised to the phase $p_i$ over $k$ periods. Eq.~\eqref{eq:corollary} underlines that the sampling in ensemble averaging can start from any period, but must be synchronised with the same starting phase and the same phase interval.

\subsection{Equivalence of ensemble averaging and phase averaging in a periodically driven turbulent flow}
The phase average $\langle\cdot\rangle$ is simply applying averaging of the same phases over all the periods in one run of the dynamic system,
\[
\langle \tilde{\bm{u}} \rangle = \int_{\bigcup^\infty_{n=1}Y_n} \tilde{\bm{u}}(\bm{r},p+nT) dP =\lim_{N_T\to \infty} \frac{1}{N_T}\sum_{n=1}^{N_T} \int_{Y_n} \tilde{\bm{u}}(\bm{r},p) dP ,
\]
where $(\bm{r},p)\in \Omega\times (0,T]$, $\tilde{\bm{u}}(\bm{r},p+nT)\in Y_n$ denotes the $n$th period of the dynamics system from the first run. The phase average measures the same phases over period in the same run of the dynamic system.

Applying Eq.~\eqref{eq:corollary} and probability conservation under period-translation Eq.~\eqref{eq:probabilityconservation}, phase averaging is equivalent to ensemble averaging, defined in Eq.~\eqref{eq:ensembleaverage}, as any periodic translation in one run of the dynamic system still retains the same probability conservation to the initial condition of the same run of the dynamic system,

\begin{eqnarray*}
\langle \tilde{\bm{u}} \rangle&=&\lim_{N_T\to \infty} \frac{1}{N_T}\sum_{n=1}^{N_T} \int_{Y_n} \tilde{\bm{u}}(\bm{r},p) dP\\
 &=&\lim_{N_T\to \infty} \frac{1}{N_T}\sum_{n=1}^{N_T} \int_{\mathcal{H}} \tilde{\bm{u}}(\bm{r},p) dP\\
 &=& \langle \tilde{\bm{u}} \rangle_e.
\end{eqnarray*}
It is also straightforward to deduce that the ensemble averaged moments equal the moments from phase averaging:
\begin{eqnarray*}
&\langle \tilde{\bm{u}}(\bm{r},p_1)\tilde{\bm{u}}(\bm{r},p_2)\cdots \tilde{\bm{u}}(\bm{r},p_m)\rangle \\
=&\langle \tilde{\bm{u}}(\bm{r},p_1)\tilde{\bm{u}}(\bm{r},p_2)\cdots \tilde{\bm{u}}(\bm{r},p_m)\rangle_e.
\end{eqnarray*}

The equivalence of ensemble averaging and phase averaging in periodically driven flows not only means that the dataset can consist of (statistically) sufficient periods in one periodic experiment, but also means that the acquisition can be implemented by several identical (but independent) experiments with synchronised phase in the acquisition process.

\subsection{In-equivalence of ensemble averaging and time averaging in periodic turbulent flows}
In periodic flows, the probability conservation condition of time averaging can not hold in general. For example, if the boundary condition of a flow is a sinusoidal function $f(t)=\sin(2\pi t/T)$ with initial phase $p_0=0$, the mean velocity of a time average near the driven boundary is zero. But the fluctuating part of the velocity in the first half period $t\in(nT,nT+1/2T]$ is positive, which means $P(u(t)<0)=0$ and $P(u(t)\ge 0)=1$. In contrast, across the second half period $t'\in(nT+1/2T,nT+T]$ the fluctuating velocity near the driven boundary is negative, which means $P(u(t')>0)=0$ and $P(u(t')\le0)=1$. This does not satisfy the probability conservation condition of time averaging $\forall t\in \mathbb{R},\, P(G_t^{-1}A)=P(A)$, because it does not hold as $t=T/2$.

In the practice of stationary flow, the experimentalist often chooses twice the integral timescale as the minimum time interval to avoid correlation between the samples, which fulfills the probability conservation condition. In periodic flows, time averaging does not equal ensemble averaging because the different phases share different probability distributions.

\section{Numerical implementation to two-dimensional flows}\label{appB}
For two-dimensional statistically non-stationary flows, the space-time domain $\Omega$ is formed by $[0,L_1]\times[0,L_2]\times(0,T]$ with $N_1$, $N_2$ and $N_p$ discrete sampling points in each coordinate.
Each velocity component $u^{\alpha}_i$ is organized in a form of three-dimensional matrix with size $N_1\times N_2\times N_p$. The velocity vector $\bm{u}^\alpha\in \mathcal{H}$ includes two velocity components $\{u^{\alpha}_1, u^{\alpha}_2 \}$. The inner product based on discrete data is defined from the definition of TKE as $(\bm{u}^\alpha,\bm{u}^\beta)=\bm{u}^\alpha_1\cdot \bm{u}^{\beta *}_1+\bm{u}^\alpha_2\cdot \bm{u}^{\beta *}_2$. The data vector $\bm{q}^\alpha = \mathrm{vec}(\bm{u}^\alpha)$ is formed by the algorithm \ref{alg:b}.

\begin{algorithm}[H]
\caption{Data vector arrangement with $N_d=2$}
\label{alg:b}
\begin{algorithmic}
\REQUIRE  $u_1^{\alpha}(\bm{x})$, $u_2^{\alpha}(\bm{x})$
\ENSURE $\bm{q}^\alpha=[\bm{q}^{\alpha}_1, \bm{q}^{\alpha}_2]^T$
\FOR{$p=1:N_p$}
    \FOR{$x_2=1:N_2$} 
        \FOR{$x_1=1:N_1$}
            \STATE $\bm{q}^{\alpha}_1(b)\gets u_1^{\alpha}(x_1,x_2,p)$
            \STATE $\bm{q}^{\alpha}_2(b)\gets u_2^{\alpha}(x_1,x_2,p)$
            \STATE $b\gets b+1$
        \ENDFOR
    \ENDFOR
\ENDFOR
\end{algorithmic}
\end{algorithm}

Here $N_1$, $N_2$ and $N_p$ are the amount of discrete sampling points in the coordinates $x_1$, $x_2$ and $p$.
The empirical velocity vector $\bm{q}^\alpha$ is a vector with $2 N_1  N_2  N_p$ rows. We take $M=2 N_1 N_2 N_p$ and total periods $N_T$ to simplify the expression of data vector, thus the empirical velocity vector $\bm{q}^\alpha$ is a vector with $M$ rows, and all the empirical velocity vectors can form a velocity matrix $\mathbf{u}$ of size $M\times N_T$. The inner product of the vector form is the same as the vector dot product, $(\bm{u}^{\alpha}, \bm{u}^{\beta}) = \bm{q}^{\beta *}\cdot \bm{q}^{\alpha}$. In this Hilbert space $\mathcal{H}$ , the correlation matrix is
\begin{equation}
\mathbf{R}=\langle \lbrace\bm{q}^{\beta}\bm{q}^{\beta*}\rbrace_{\beta=1}^{N_T}\rangle =
\begin{bmatrix}
\mathbf{R}_{11} & \mathbf{R}_{12} \\
\mathbf{R}_{21} & \mathbf{R}_{22} \\
\end{bmatrix}.
\end{equation}

\section{Convergence of statistically non-stationary lid-driven cavity flow}\label{appC}
\begin{figure*}[ht]

        \begin{subfigure}[b]{0.45\textwidth}
        \includegraphics[width=\textwidth]{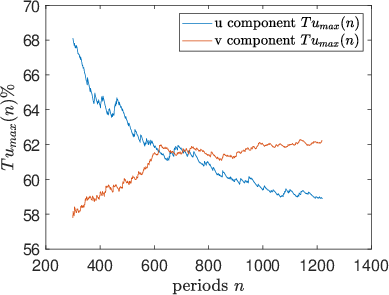}
        \caption{\label{fig:statistics1}}
        \end{subfigure}
        \hfill
        \begin{subfigure}[b]{0.45\textwidth}
        \includegraphics[width=\textwidth]{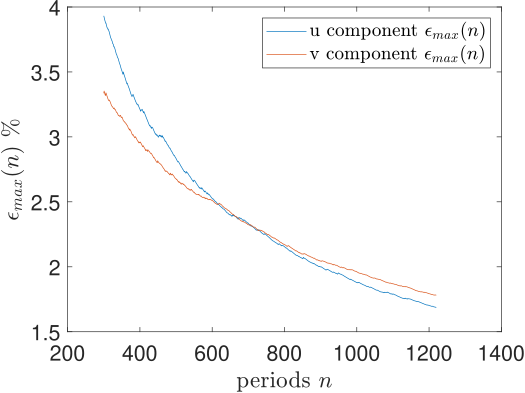}
        \caption{\label{fig:statistics2}}
        \end{subfigure}
        
    \caption{Convergence of the statistical non-stationary lid-driven cavity. \protect\subref{fig:statistics1} Turbulence intensity $Tu_{max}(n)$ of both $u$ and $v$ components. \protect\subref{fig:statistics2} Variability $\epsilon_{max}(n)$ of both $u$ and $v$ components.}
\end{figure*}

Statistical evaluation of the turbulent non-stationary lid-driven cavity flow is based on the turbulence intensity $Tu(\bm{x},n)$ and variability of mean velocity $\epsilon(\bm{x},n)$, which is the estimated normalized sampling error in Oliver et al.\cite{Oliver2014} with the decorrelation separation distance being equal to unity. To check the convergence of the data, we take different sample amounts of the data to observe the convergence. The turbulence intensity is defined as (see p. 311 in Ref.~\onlinecite{George2013}):
\[
    Tu(\bm{x},n)=\frac{\sigma( \bm{x}, n)}{\langle \tilde{u}( \bm{x}) \rangle},
\]
where $\sigma( \bm{x}, n)$ is the standard deviation of samples from 1 to $n$. The variability of the mean velocity $\epsilon(\bm{x},n)$ is defined as
\[
    \epsilon^2(\bm{x},n)=\frac{(\frac{1}{n}\sum_1^n \tilde{u}( \bm{x})-\langle \tilde{u}( \bm{x}) \rangle)^2}{(\langle \tilde{u} (\bm{x}) \rangle^2}.
\]
The variability of the mean velocity and the turbulence intensity are directly related to the (independent) sample number $n$ as
\[
\epsilon(\bm{x},n)=\frac{Tu(\bm{x},n)}{\sqrt{n}}.
\]

Because $\epsilon(\bm{x},n)$ and $Tu(\bm{x},n)$ are four-dimensional data, the maximum turbulence intensity and variability are employed to check the convergence of the random velocity vectors (see Theorem 5.4.1 in Refs.~\onlinecite{Hogg2019}):
\[
Tu_{max}(n)=\max_{\substack{\bm{x}\in \hat{\Omega}}}\{Tu(\bm{x},n)\}.
\]
\[
\epsilon_{max}(n)=\max_{\substack{\bm{x}\in \hat{\Omega}}}\{\epsilon(\bm{x},n)\}.
\]

To avoid singular values where $\vert\langle u( \bm{x}) \rangle\vert\approx0$, we pick a subset of data where the mean velocity is 1 000 times above the machine error. The results are plotted in Fig.~\ref{fig:statistics1} and \ref{fig:statistics2}.

The turbulence intensity is stable at around 59\% for the $u$ component and at around 62\% for the $v$ component after 900 periods. At the same time, the maximum variability of both the $u$ component and the $v$ component is lower than 2\%, which falls within an acceptable range of demonstrating the Phase POD method.

\begin{figure*}[ht]
     \centering
        \includegraphics[angle=90,origin=c,width=0.8\textwidth]{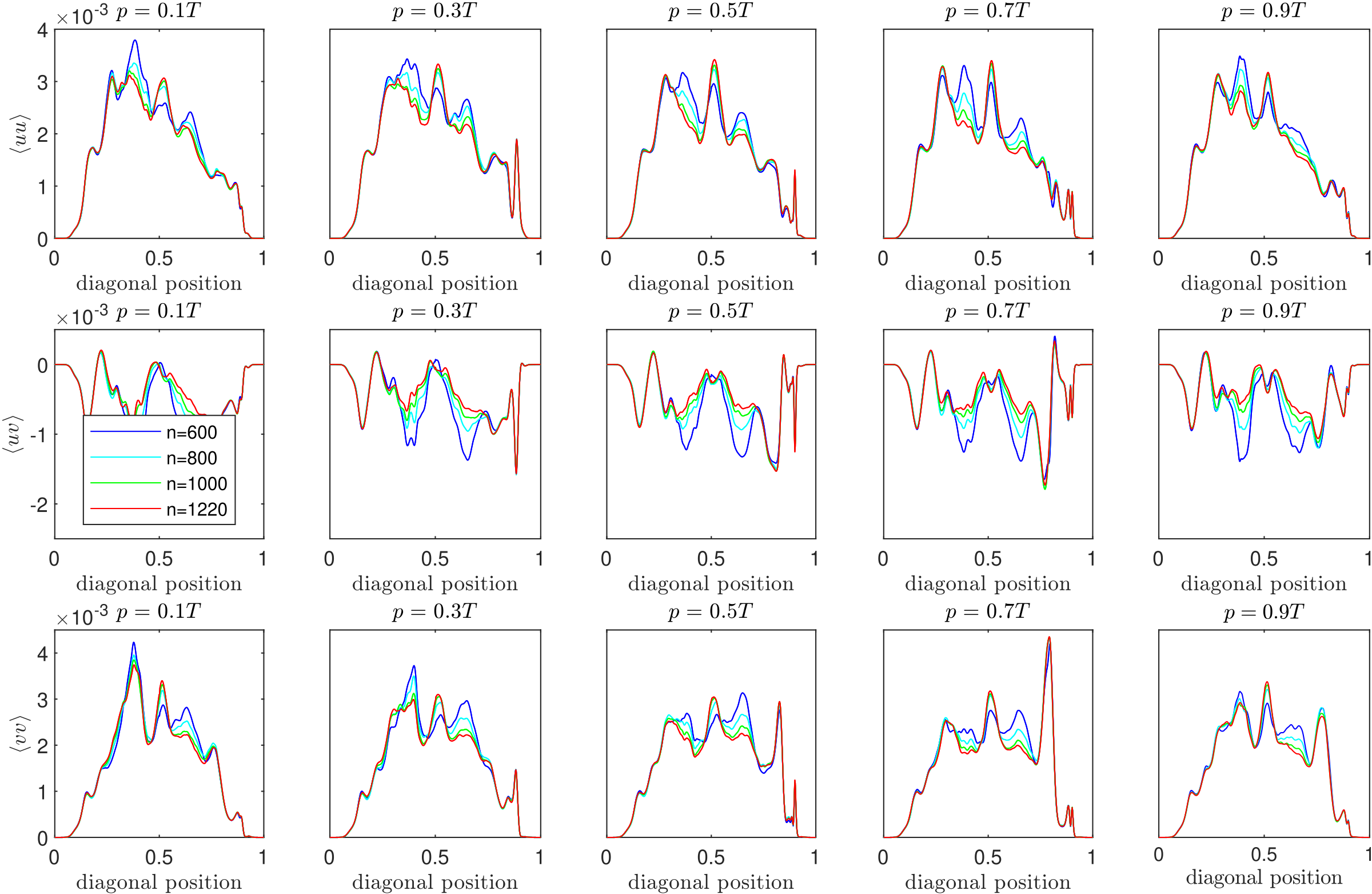}
        \caption{Reynolds stress of all three components $\langle uu \rangle$, $\langle uv \rangle$ and $\langle vv \rangle$ at the diagonal point in the cavity over some phases. The Reynolds stresses of all diagonal points and phase are converging as $n$ increases.}
        \label{fig:statistics}
\end{figure*}

\begin{figure*}[ht]
     \centering
        \includegraphics[angle=90,origin=c,width=0.8\textwidth]{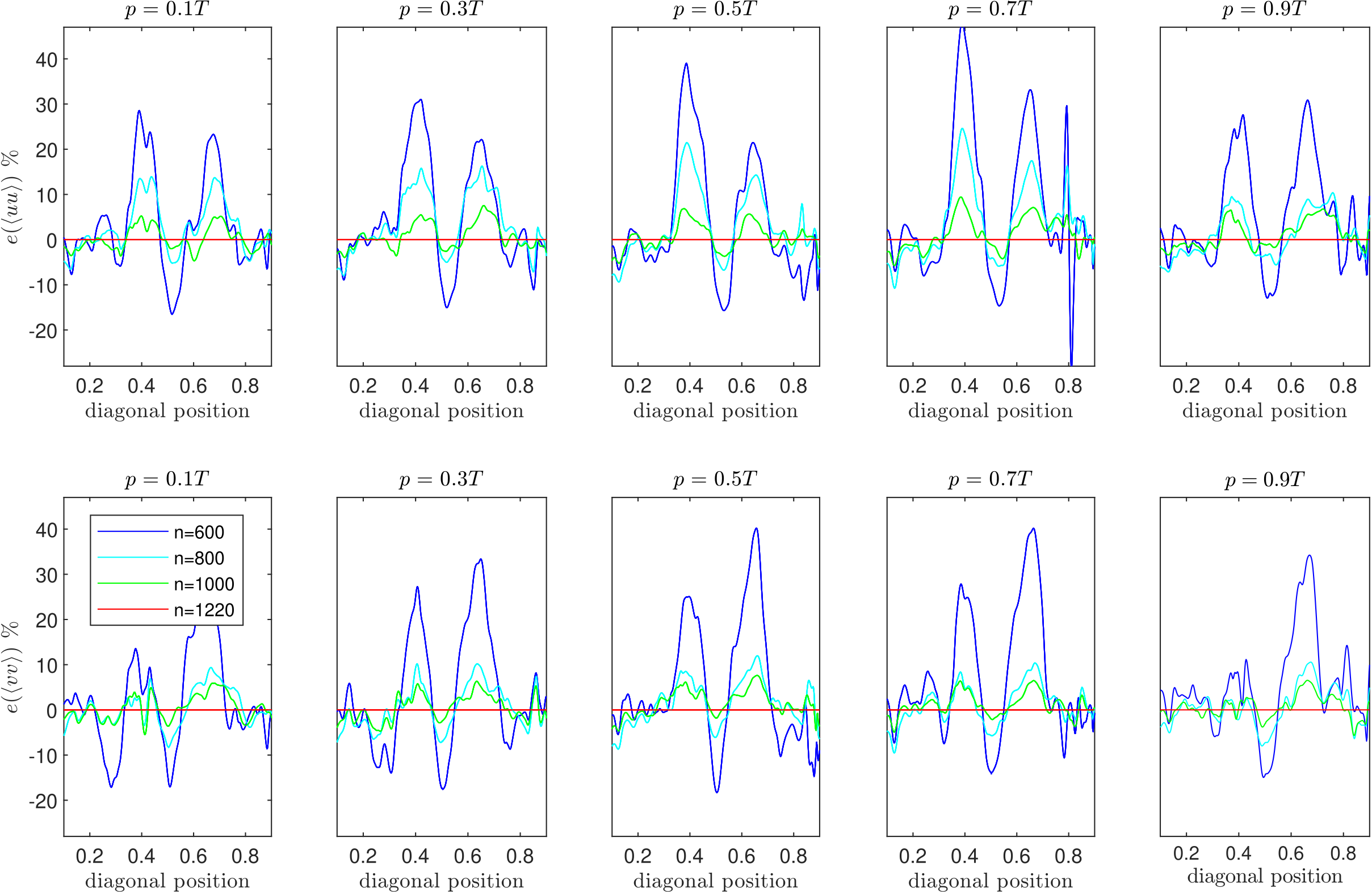}
        \caption{Reynolds stress errors of $\langle uu \rangle$ and $\langle vv \rangle$ at the diagonal point in the cavity over some phases. The Reynolds stresses of all diagonal points and phase are converging as $n$ increases.}
        \label{fig:statisticserros}
\end{figure*}

The Reynolds stress tensor is decomposed into POD modes. In Fig.~\ref{fig:statistics}, we compute all three components $\langle uu \rangle$, $\langle uv \rangle$, and $\langle vv \rangle$ on the diagonal of the cavity over all phases. To visualise the errors with increasing $n$, Fig.~\ref{fig:statisticserros} shows the corresponding Reynolds stress variability of $\langle uu \rangle$ and $\langle vv \rangle$ at the diagonal point in the cavity over some phases. It is easy to observe that the Reynolds stress is converging statistically with increasing $n$. 

\nocite{*}
\bibliography{aipsamp}

\end{document}